# Towards Harmonious Decentralization of Energy Systems: A Vision of Interoperable Peer-to-Peer Energy Markets

by

Sakshi Mishra, Roohallah Khatami, Yu Christine Chen

Department of Electrical and Computer Engineering

The University of British Columbia, Vancouver, BC, Canada

# Abstract

We present a hierarchical framework aimed at decentralizing the distribution systems market operations using localized peer-to-peer energy markets. A hierarchically designed decision-making algorithm approaches the power systems market operations from a bottom-up perspective. The three layers of the hierarchical framework operate in orchestration to enable prosumers (the grass-root ac- tors) to maximize their revenues - hence, a prosumer-centric framework. The design of the framework incorporates existing smart grid technologies (Virtual Power Plants, Microgrids, Distributed Energy Resources) and redefines their functional objectives to align them with the decentralization paradigm focused on empowering the bottom-up grid operations approach. On one hand, the framework is enabling prosumers with simultaneous access to the buy-sell choices that help them maximize their cost savings while ensuring their consumption patterns and preferences are not being traded off as a result of top-down operational decisions. On the other hand, it is designed to operate in harmony with the existing top-down grid operations mechanisms - thereby reducing the potential friction in its adaptation. This marriage of the top-down and bottom-up operational approaches is facilitated through meticulous orchestration of operational timescales. The framework's novel design also incorporates scalability and interoperability considerations, thereby tackling the challenge of decentralization holistically.

# 1 Introduction

## 1.1 Motivation

Energy systems are part of the critical infrastructure of a nation. Electricity not only keeps the lights on for the end-users but also forms the backbone of industrial activities. In today's digital age, where electronic devices are cornerstone of modern life, the importance of reliable electricity supply – the fuel for these devices – has only escalated many folds. At the same time, energy systems are amidst a major transformation. The unprecedented changes in the global energy landscape are driven by a number of factors (listed in Appendix I).

The electricity sector has primarily been monopolistic dominated by vertically integrated utilities. Liberalization of electricity markets over the past several decades has opened doors *only* for owners of utility-scale generation plants to participate. From the *end user's perspective*, the electricity sector remains largely centralized: end users have very little (if at all) say in market dynamics and therefore they have no choice but to be "price-takers". The proliferation of DERs, reaching high penetrations, offers a beacon of hope: it empowers the end users and their communities in unprecedented ways toward self-sufficiency to the extent that going "off-grid" is starting to become a viable option. Depending on the jurisdiction, prosumers[1] may already have the option to sign contracts with *DER aggregators*–third party *for-profit* entities serving as intermediaries between the prosumers and wholesale markets. Such contracts often lock prosumers into selling their electricity for a stipulated length of time (often a few years) and therefore do not necessarily maximize the value of energy that prosumers offer. And so the mechanism of such contracts is not effectively serving the purpose of decentralized operations where prosumers are the primary decision-makers.

We, therefore, re-imagine the operational paradigm of decentralized energy systems by modeling the bottom-up approach to grid management. Moreover, in this work, we put forth the idea that the centralized (top-down) vs. decentralized (bottom-up) architectures for grid operations are not necessarily at fundamental odds.

> We propose that essential to a paradigm shift in the electricity sector's operating philosophy is to find synergies between top-down and bottom-up approaches of managing the generation, transmission, distribution, and end-use pipeline.

In practical terms, this proposition entails that the prosumers or the end users are no longer passive price-takers. Instead, they participate in the wholesale markets as price-makers. At first glance, the idea of top-down and bottom-up approaches coexisting may seem paradoxical, and we may be tempted to label the bottom-up approach of infusing the grid edge with greater control coupled with DERs deployments as "disruptive". However, we must be mindful of the fact that technological disruptions in the electricity sector cannot and will not happen overnight given the conservative nature and sheer scale of this industry[2]. Furthermore, the past decade has borne witness to significant [1] [2] [3] pushback from utilities entrenched in the top-down architecture against DERs and community microgrids as these concepts challenge existing business models. If the bottom-up operational approach leans toward augmenting the objectives of the current top-down approach, rather than outright displacing the centralized system, it will more likely gain widespread adoption. The marriage of top-down and bottom-up architectures, nonetheless, will require novel coordination mechanisms that can aggregate small-scale DERs toward system-wide objectives while fully preserving their owners' energy choices (i.e., when and how they would like to use the energy from their DERs).

---

[1] Prosumers are "proactive consumers" who own on-site energy production and storage capabilities. They undertake a proactive behavior by managing their consumption, production through DERs, and energy storage.
[2] In 2022, the electric power industry in the United States generated a revenue of about 488.4 billion U.S. dollars [32].

Meanwhile, it is also essential to identify different value streams for revenue generation to incentivize end-users (i.e. prosumers) to continue investing in DERs deployment projects. In theory, a straightforward value-stream creation option is to connect these prosumers to the already existing *wholesale markets* for energy, transmission rights, and ancillary services. However, this proposition falls short of practical value because of a number of barriers inhibiting DERs' participation. "Individual consumers and small-scale generators have a little individual impact at the transmission level, and the complexity of the processing and communications infrastructure needed for participation would involve transaction costs that outweigh potential benefits. Furthermore, market price fluctuations present risks that are difficult to hedge at the individual level. Instead, they have traditionally been serviced in retail markets by large suppliers, where economies of scale allow for transaction costs to be overcome and market price risks to be diversified." [4]. However, this barrier to entry into the wholesale markets renders prosumers in a state where they only benefit from their flexibility to shift their own peak demand resulting in reduced demand charges.

*Coordinated local DERs* can overcome some of these challenges effectively by offering win-win solutions for grid operators and end-users or prosumers. Firstly, when operating as an orchestrated fleet of generation and storage technologies, DERs are able to participate in the wholesale markets - opening up a value stream for the individual owners. In the process of participating as a fleet, they can offer the grid operators an opportunity to harness their services (for example, in the ancillary market) without having to undertake the burdensome task of monitoring & controlling individual small scale generation technologies[3].

Secondly, they can reduce the upstream generation and transmission capacity requirements by meeting large proportions of the demand within the coordinated cluster from the local generation itself. Note that the capacity build-out by the utilities is a favorable outcome as far as the existing utility business models are concerned. However, several system-wide challenges exist for building out new grid infrastructure: i) Siting new transmission lines (getting approval of new routes and obtaining rights to the necessary land); ii) Determining an equitable approach for recovering the construction costs of a new transmission line built in one state when the line provides benefits to consumers in other states; iii) Expanding the network of long-distance transmission lines to renewable energy generation sites where high-quality wind and solar resources are located, which are often far from where electricity demand is concentrated; and so on. Therefore, a reduction in the upstream capacity requirements is a favorable outcome from both regulator's and the prosumer's perspective.

Thirdly, the coordinated DERs can help achieve cost savings on utility bills (by peak demand reduction) and provide resilience benefits such as sustaining the critical load during the utility outage. The cost-savings are achieved when the peak demand of the individual prosumer is met by the local generation from the coordinated fleet, thereby reducing the utility *demand charges* imposed on individually metered prosumer. These cost-savings are also reflective of the energy efficiency that the grid is able to achieve by not having to move the power over long distances and having Peaker plants turned on to meet the demand of the consumers during on-peak hours. Furthermore, these DERs can help enhance the resilience of the *distribution grid* by providing black-start services in the event of a blackout.

Given the myriad of benefits associated with operating the DERs as a coordinated fleet, this concept has gained momentum recently and has been termed as **Virtual Power Plants** (VPPs). A VPP is a collection of DERs that are coordinated to have visibility, controllability, and impact at the transmission level of the power network [5]. VPPs are discussed further in the background [section 2].

On the other hand, for individual prosumers, there exists one more option to get remunerated for the excess energy their DERs generate, it is called the 'net-metering' policy. In this arrangement, prosumers essentially sell their electricity back to the grid at a "buy-back rate". However, the price of electricity that they buy from the grid is generally much higher than the buy-back rates that prosumers can obtain from selling the electricity to the grid. Therefore, a new type of revenue generation instrument has been emerging for the prosumers. This approach or the instrument is termed as *peer-to-peer* (P2P) *energy markets*.

P2P energy markets capacitate the prosumers to trade electricity in the absence of intermediaries at their agreed price, by establishing a platform to transact with each other. It is an appealing option that becomes increasingly viable with greater DER penetration. This market apparatus further spurs decentralization by empowering prosumers to have greater control over the use of energy generated by the DERs they own. Thus, P2P energy markets will play a significant role in propelling the bottom-up approach to energy systems operations. Moreover, when designed appropriately, P2P energy markets not only empower the prosumers but also offer many potential benefits to the central grid such as aiding with congestion management and providing ancillary services.

Although capable of functioning as a separate mechanism, peer-to-peer energy markets are largely going to operate amongst prosumers who are connected to the utility grid[4]. And in this grid- connected setup of P2P energy markets, prosumers can also get benefited by offering their DER generation to the centralized markets - hence the coordinated fleets (VPP discussion in the previous paragraphs). Therefore, it is imperative that combining P2P energy markets with the VPPs is an optimal way for opening maximum revenue generation streams for the prosumers, thereby nurturing the decentralization paradigm.

## 1.2 Problem Statement

For more than a century, the grid has operated in a centralized top-down fashion. However, as the penetration of distributed energy resources (DERs) grows, the grid edge is increasingly infused with intelligent computing and communication capabilities. Thus, the bottom-up approach to grid operations inclined toward decentralization of energy systems will likely gain momentum alongside the existing centralized paradigm. Indeed, recent academic research has focused on local decentralized energy trading [6] [7] [8] [9] [10] [11] [12] There are also test-bed and prototype deployments around the world to demonstrate the feasibility of localized peer-to-peer energy trading (see [13] for a summary) and microgrid initiatives by government [14]. This growing trend toward the bottom-up approach will challenge the institution of the traditional top- down architecture for grid operations. Easily anticipated are frictions that will arise amongst different actors along the electricity supply chain and in electricity markets (e.g., generation owners, large utilities, small DER owners, policymakers, etc.) as decentralization escalates. Therefore, critical in a credible and practical path forward is to coordinate the coexistence of the two seemingly contradictory approaches of grid management, toward the overarching goal of reforming the grid of the future as a reliable, efficient, decarbonized, and economical infrastructure to meet the growing global energy demand.

---

[3] In transmission level power systems models, the loads are often considered in bulk. In other words, loads are represented in an aggregated manner. Consequently, the granularity of the load modeling is lost on large-scale grid models. Operators, therefore, have no visibility into the small-scale DERs despite these devices being equipped with modern sensors and controllers

[4] It is possible for a group of prosumers to go off-grid altogether when the DER systems are designed with the capacity to sustain all the load. However, our focus in this research work is on prosumers who operate in a grid- connected fashion while also choosing to participate in the peer-to-peer energy markets.

Given the pace of transformation the energy landscape is undergoing, there are multiple niches where the research frontiers are being pushed further from generation to transmission to distribution to end-user domains. Innovations in these niches, which touch upon and at times encompass the technical, economic, and policy aspects of energy systems, are very much needed to help meet the decarbonization targets and sustainability goals [15]. There exists, however, a dire need to bring together the advances in different niche areas and understand their workings in *a piece of a whole* context and with that awareness shape cognizant decentralized operations of the energy systems that gel well with existing top-down operations. By doing so, we open up the pathways of enhancing the functionality of these standalone smart grid technologies to better serve the objective of creating a clean and equitable energy future for the planet, by utilizing the feedback loop originating in this *systems approach*. In this process, the existing research on these niche areas becomes more and more practically relevant. Consequently, the research-to-deployment timeline of these cutting-edge concepts and ideas will be shortened. Furthermore, it is vital to find a way to make these smart grid technologies and operative paradigms work together so that their synergies can be channelized for achieving the higher-level objectives of reducing carbon footprint, maintaining reliability, and increasing the overall energy efficiency of the electricity supply chain.

This research, therefore, is focused on taking a holistic approach to architect a **prosumer-centric grid operations framework**, accounting for the multi-faceted intricacies of power systems markets and operations, that can help bridge the gap between the top-down and bottom-up approaches of electricity grid operations management on the distribution grid end. On one hand, this approach empowers the prosumers with greater control over the energy produced by the DERs they own. On the other hand, it gels well with the existing large-scale centralized markets.

## 1.3 Novelty and Contributions

Works in the literature on P2P energy trading do not consider P2P trading in conjunction with decentralized VPPs. For example, Paudel et al. [16] study P2P energy trading with network utilization costs while Ullah et al. study P2P energy trading with voltage management constraints [17]. Other studies such as from Singh et al. [18] and Azim et al. [19] only focus on P2P trading amongst the peers and not the interaction with the wholesale markets. Moreover, the P2P trading mechanisms proposed in the literature above do not consider the real-life regulatory structures that enable or prohibit the interactions of retail customers with the wholesale markets. In this work, we have built the foundations of the proposed framework on two crucial regulatory and cost-allocation factors:

1. FERC ruling (Order No. 2222): This order has, for the first time in the U.S., opened the "regulatory" doors for small-scale DER participation in the lucrative wholesale markets.
2. Redefining the functional role and ownership aspects of Virtual Power Plant (VPP) entities such that they do not act as for-profit third-party intermediaries or aggregators that: A) Take a cut from prosumers earnings from seeling prosumers' DERs energy in the wholesale market and B) Limit the direct options that prosumers have available for selling their production at any given point in time (aka simultaneous choices to sell prosumer energy in local P2P markets, retail behind-the-meter market, and wholesale market).

To realize the vision of developing this prosumer-centric P2P energy trading framework, designed to work alongside the top-down approach to grid management, we have mapped out and accomplished the following specific objectives:

- Establish the value-proposition of decentralized energy systems and put forward the advanced notion of operational planning within the decentralization context.
- Prototype the hierarchical architecture of decentralized Peer-to-Peer markets by acknowledging the preordained interactions between *top-down* and *bottom-up* approaches to electricity grid operations.
- Model the ladder of decentralized dispatch decisions in the hierarchical P2P markets - considering the complex interplay of the physical, market, and cyber layers.

The resistance posed by the incumbent utilities and traditional grid operational mechanisms in the regulatory environment, has made it practically impossible for prosumers to participate in a fair manner. This framework, in its essence, offers a fundamental fabric for enabling the co-existence of the centralized and decentralized electricity market paradigms. Thereby, providing a fertile ground for the emerging decentralization paradigm to thrive to contribute to the sustainable and low-carbon future of the planet. It is worth noting that our proposed framework places high value on "decentralization paradigm" where prosumers truly can *have* and also exercise their power to sell their energy to maximize their savings – a hallmark of true decentralization of energy markets. At the same time, our framework doesn't oppose or suggest to altogether demolish the top-down markets, It instead finds an innovative way to connect both with multi-layers operational structures.

## 1.4 Article Structure

The motivation, problem statement, and contributions of this work were presented in the current section (section I). Section II explains the building blocks of the framework (smart grid technologies and concepts) with their associated functionalities. It also discusses the prior art in VPPs and the P2P energy trading domain. Section III presents the end-to-end design of the proposed framework and delineates the operational scheme tailored for harmonious interactions of the top-down and bottom-up approaches to grid operations. The design of the framework is hierarchical in nature with three distinct layers within the market sphere, constituting a highly conducive environment for a prosumer-centric decentralized operational paradigm to exist. Section IV concludes the article and discusses future research directions.

# 2 Background

## 2.1 Connections Between Existing Centralized Energy Markets and Proposed Peer-to-Peer Markets

The range of benefits that decentralized and peer-to-peer energy markets offer , to the prosumers primarily located on the distribution grid, can only be utilized in its entirety when analyzed in conjunction with their interaction with the central grid's market operations. This is because the *transmission* grid & *distribution* grid operations are growing more and more intertwined because of the bidirectional flow of power. Furthermore, the coordinated fleet of DERs offers benefits such as demand response and load flexibility to the grid operators and at the same time fosters an additional value stream for the prosumers. This mutually beneficial proposition asserts the need for considering the interactions between the centralized (wholesale and retail) markets and localized coordinated DERs operations (i.e., VPPs).

The majority of the centralized electricity markets have two components - retail and wholesale. While retail markets involve the sales of electricity to consumers, wholesale markets typically involve the sale of electricity among electric utilities and energy traders before it is eventually sold to the consumer through the retail markets[5]. Operations of the wholesale and retail markets involve a number of participating agents in the power systems. These agents include utilities, grid operators, load-serving entities, third-party service providers (in retail markets), energy traders, and so on. A set of complex interactions between these agents govern the market mechanism that supplies electricity.
to the end users. A brief description of the agents and their roles is provided in Appendix II.

For a P2P energy market to be effective in practice, it is important to position its design such that it takes care of the agent interaction dynamics in the electricity supply chain. Therefore, the following subsections are dedicated to describing the concepts surrounding wholesale and retail markets and their relation to the physical grid operations which constitute the backdrop for designing the proposed decentralized peer-to-peer energy markets framework.

### 2.1.1 Retail Deregulation

In retail deregulated areas, electricity customers have the option of selecting an electric supplier (known as customer choice) rather than being required to purchase electricity from their local electric utility, which introduces competition for retail electricity prices. Since many electric suppliers can exist within a region with customer choice, electric retailers offer competitive prices in order to acquire customers (contracts with generation suppliers typically offer the customer a fixed charge dollars per kilowatt-hour of power over a certain amount of time).

For consumers, there are pros and cons to selecting a supplier other than their local utility company. Retail competition can help lower a customer's electric bill and also allow them to tailor their energy to their preferences, such as selecting a clean energy supplier. However, **independent companies** often require customers to sign contracts, which can lock them into a set electricity price for multiple years. While fixed rates could be beneficial for some customers, they could also negatively impact others if the rate they agree to ends up being more expensive than the rate set by the local utility. Also, it is important to note that customer choice is only applicable for the generation portion of a customer's utility bill because transmission and distribution services are still provided by the local utility company since these services are a natural monopoly (as discussed above). Consequently, only a portion of electric rates in these areas are set competitively. For customers who choose not to select an independent power supplier, their local utility is still obligated to provide them with electricity that the utility will purchase from generators.

---

[5] Not all the states or provinces fall neatly into the category of having both retail and wholesale electricity markets. Some states have deregulated their wholesale markets but not retail markets. Moreover, in the competitive retail markets, municipally-owned utilities may not offer their customers retail choices. It is important to note that the market is not always divided clearly between traditionally regulated markets and competitive market states. Some states, like California, are partially restructured markets and only permit certain consumers to engage in retail choice.

> In our proposed framework, we attempt to position the functioning of VPPs as a replacement for the independent companies mentioned in the previous paragraph. Architecting the operations of VPPs such that they are not acting as a *for-profit centralized aggregator* and instead operating *semi-autonomously* with the objective of maximizing prosumer savings is part of the challenge we undertake in designing the framework.

In the long term, these VPPs should be capable of interacting with both retail and wholesale markets. However, in our line-of-research for this work, we choose to focus our framework's interaction with wholesale markets. The primary reason behind this choice is the prevalence of wholesale markets compared to the retail markets in the Northern American power grid. Another reason behind studying the interactions of the proposed framework with the wholesale market is that the goal is to help prosumers derive the most value out of their DER's energy production, and wholesale markets (especially the markets that run on a short time frame) are often lucrative.

### 2.1.2 Wholesale Markets Structure

In the deregulated regions, grid operators, also called RTOs[6] and ISOs[7], typically run three kinds of markets that determine wholesale prices for selling electricity: i) Energy Markets; ii) Capacity Markets; and iii) Ancillary Markets. A brief description of these three types of markets is provided in Appendix III.

The three sub-types of the wholesale markets run by ISO/RTOs, working on different time scales to maintain the grid reliability, are shown in Figure 1.

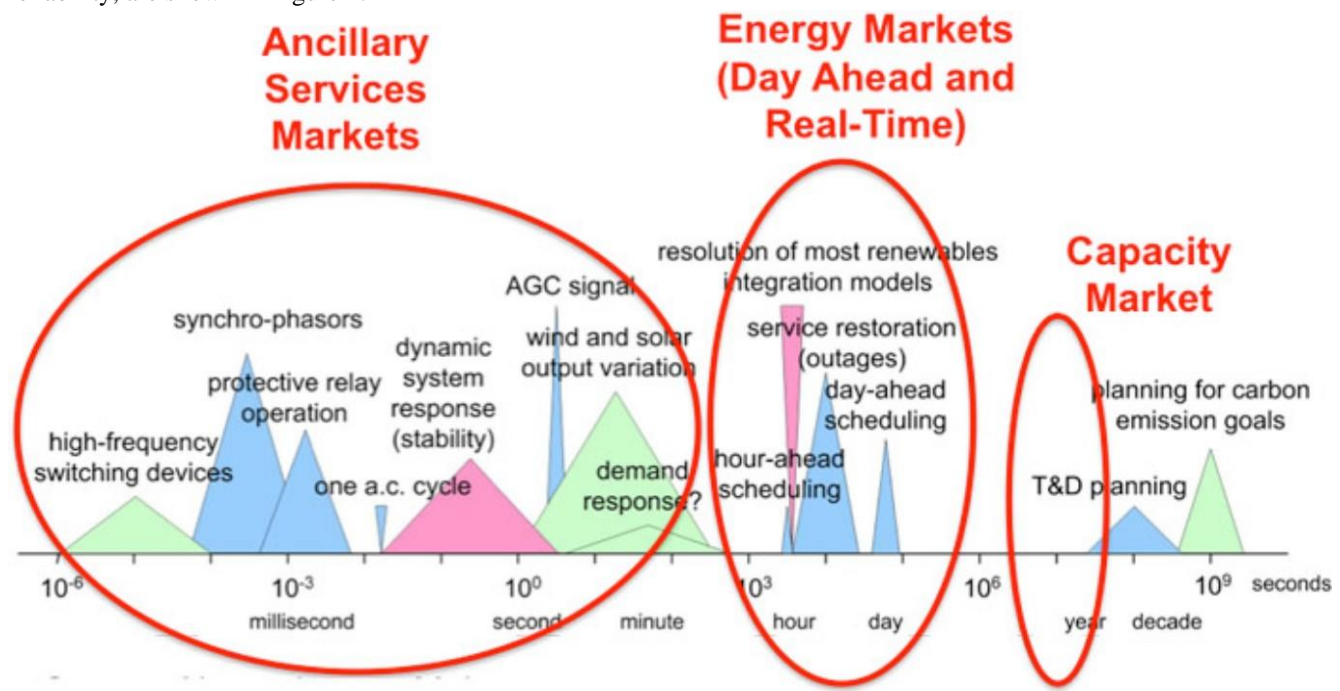

*Figure 1 Market constructs used by the ISO/RTO to ensure the reliability of electricity supply over multiple time scales.*

## 2.2 Decentralized Peer-to-Peer Energy Markets: Technologies at Play

The proposed hierarchical framework (described in section 3), in part, offers an effective way of harnessing existing energy technologies (like VPPs, microgrids, and P2P energy trading) to achieve the objective of scalable and interoperable decentralized distribution grid level P2P energy markets leading to greater autonomy, higher efficiency, and increased resilience of the localized generation. For some of these technologies, there are slightly varying interpretations of their functionalities and roles in literature. Therefore, in the following subsections, we elucidate on the underlying functionalities and the capabilities of these technical apparatuses as they are the vital constituents of the proposed framework.

### 2.2.1 Distributed Energy Resources

Distributed Energy Resources (DERs) are small or medium-sized resources, directly connected to the distribution network [20].

---

[6] Established pursuant to Federal Energy Regulatory Commission (FERC) Order No. 2000, an RTO is a federally regulated independent entity that is responsible for: i) Managing, providing access to, and monitoring transmission facilities under its control; ii) Maintaining grid stability and reliability; iii) Matching electricity demand to supply; iv) Facilitating competition among wholesale electricity suppliers.

[7] An independent and federally regulated entity that coordinates regional transmission to ensure non-discriminatory access to the electric grid and a reliable electricity system. ISOs were formed following the issuance of FERC Orders Nos. 888 and 889 by the Federal Energy Regulatory Commission (FERC) to encourage utilities to work together to ensure they provided non-discriminatory access to their transmission lines to non-utility-owned power plants. ISOs are not the only transmission operators in the US. There are also regional transmission organizations (RTOs) that perform generally the same functions as ISOs, but RTOs have jurisdiction over a larger geographic area.

In the U.S., FERC Order No. 2222 expansively defined DERs as "any resource located on the distributed system, any subsystem thereof or behind a customer meter." [21]. DERs include electric storage resources, distributed generation, thermal storage, heat pump, flexible and controllable loads, energy efficiency measures, and electric vehicles and their supply equipment. In most instances, these resources are located close to the end user of electricity. *Demand Response* is also a form of DER.

Net energy metering (NEM) metering is a system in which DERs are connected *behind-the-meter*[8] to a distribution system, and any surplus power generated by the prosumer-owned DERs is transferred onto the grid, allowing customers to offset the cost of power drawn from a distribution utility. Such surplus flow typically occurs during periods when the DERs production outstrips the customers total demand. As a result of NEM, which is a highly conducive policy toward DERs, the overall quantity of DER installations has grown a lot in some states of the U.S.

The recent FERC ruling (Order No. 2222) [21] has, opportunely, enabled DERs to participate in the wholesale markets. This rule enables DERs to participate alongside traditional resources in the regional organized wholesale markets through aggregations, opening U.S. organized wholesale markets to new sources of energy and grid services. It allows several sources of distributed electricity to *aggregate* in order to satisfy minimum size and performance requirements that each may not be able to meet individually. The key here is the *aggregation*. This opens up doors for third-party service providers (who are different from Load Serving Entity responsible for serving the retail customers in retail regions). Mechanisms of aggregation, that promote scalability and interoperability, are one of the focus points in our proposed framework. We will cover this topic in-depth in the subsequent sections.

## 2.2.2 Microgrids

The definition of a microgrid is continually evolving as new concepts and technologies are introduced to the smart grid and urban planning domain. As per the U.S. Department of Energy, "a group of interconnected loads and distributed energy resources within clearly defined electrical boundaries that acts as a single controllable entity with respect to the grid. A microgrid can connect and disconnect from the grid to enable it to operate in both grid-connected or island-mode [14].. A European perspective defines a microgrid as "a low voltage distribution network comprising various DG [Distributed Generators], storage devices, and controllable loads that can operate interconnected or isolated from the main distribution grids [22]. Microgrids are fundamentally localized grids that can disconnect from the traditional grid to operate autonomously. Because they are able to operate while the main grid is down, microgrids can strengthen grid resilience and help mitigate grid disturbances as well as function as a grid resource for faster system response and recovery.

Although there seems to be no standardized classification in the academic literature or industry for categorizing microgrids based on the scale and sophistication of the assets, operations, and control functionalities, there are indeed commonalities or overlaps between various definitions. The key aspects or features of microgrids covered in the definitions include (i) the ability to operate in grid-connected and islanded modes; (ii) clearly defined electrical boundaries; and (iii) the ability to manage their own operation by controlling interconnected loads and generation resources.

Microgrids are typically connected to the utility at the campus-wide main metering point and operate without much interaction with the distribution system, except for the bidirectional power flow when excess energy is sold back to the grid. Net metering is the common mechanism used to account for the bidirectional power flow between the microgrid and the utility grid, simplifying the compensation structure. The ownership structure of such microgrids is often straightforward as well, where utilities consider them a behind-the-meter resource.

There are two main types of microgrid architectures: i) permanent islands or self-contained, and ii) grid-connected. The self-contained microgrids, often constructed in remote or completely islanded locations, have no ties with the utility grid. Consequently, their generation technologies and other infrastructure components (such as power electronics, energy storage, and fuel storage tanks) are sized to meet the entire load all the time independently. On the other hand, the grid-connected microgrids are designed to stay connected to the central grid under normal operations. They are, however, equipped with necessary controls and switch functionalities to isolate from the utility grid in the event of a utility grid outage. Grid-connected microgrids are often sized to power only a subset of the critical loads in case of utility grid outage rather than the full site-wide load, as they are usually operating in the grid-connected mode.

## 2.2.3 Virtual Power Plants

A Virtual Power Plant or VPP is a network of independent DER systems utilizing a cloud-based control system to perform like a single large-capacity energy source. It is basically a *distributed* power plant formed by aggregating the capacities of heterogeneous DERs. It performs the task of coordinating multiple distributed DERs for enabling their participation in the wholesale markets. DERs can be of a variety of types such as dispatchable and non-dispatchable, controllable or flexible loads, photovoltaics (PVs), energy storage systems, backup generators, biomass, small-scale wind turbines, etc.

VPPs have started thrusting ahead with real-world deployments. For example, in 2019, a VPP project was announced by the South Australian government (in collaboration with Tesla) for 50,000 houses [23]. The prevailing consensus on VPP's objectives is to enable the participation of small-scale DERs in the wholesale markets. The markets often have strict requirements regarding the availability and reliability of the flexibility offered in the market. VPPs enable the power and flexibility of heterogeneous assets (consisting of power generators and flexible loads) to be traded collectively in lucrative sections of wholesale markets (balancing reserves, for example). Therefore, the design of VPPs with this particular objective is such that the decentralized units

---

[8] *behind-the-meter* refers to anything that happens on site, on the energy user's side of the meter. Conversely, anything that happens on the grid side is deemed to be in front of the meter.

are linked and operated by a single, centralized control system.

Interestingly, the VPPs in the current scenario are often commissioned by the utilities or the government, along with the capital expenses for setting up the needed digital infrastructure. Though the data back-and-forth between the VPP's control system and individual assets is facilitated through public communication infrastructure, the remaining digital components such as the VPP's controls software, measurement, and control sensors on the individual assets still need capital investment. **This arrangement makes these VPPs a part of the *top-down approach* of managing the power systems**. In our research, we challenge this notion and set out to design our framework such that along with enabling the participation of DERs in wholesale markets, VPPs are acting as facilitators for localized and decentralized peer-to-peer energy markets. In other words, VPP's operational objective is not limited to maximizing prosumer savings by representing them in the wholesale markets, VPP also interfaces with the localized peer-to-peer markets. This way, VPPs act as a facilitator between the bottom-up and top-down approaches, primarily aiming to aid prosumer- centric decision-making. We discuss this further in the 3 section.

### 2.2.4 Difference between VPP and Microgrid

Both VPP and microgrid often involve a mix of DERs including storage and flexible loads, but there are some important differences that set these two mediums of DER aggregation apart:

- VPPs do not have any physical assets of their own, they more or less operate in the market and cyber layers that together do the aggregation and efficient dispatch of DERs (from different owners) that it bundles. Whereas, whereas microgrids are capable of grid-connected and islanded modes of operations. In the islanded mode, they can continue functioning on their own.
- VPPs can be assembled using assets connected to any part of the grid, whereas microgrids are usually restricted to a particular location, such as an island or a campus.
- From grid operator's perspective, VPPs are *front-of-the-meter* power plants. That is, different resources within the VPPs are metered individually. Whereas, microgrids are often considered a *behind-the-meter* (BTM) resource where all the DER assets of the microgrid are metered at the single point of interconnection to the main grid.

In our proposed framework, we further establish the difference between the *operational objectives* and *functionalities* of these two mediums of aggregation to better serve the overall objective of designing *prosumer-centric* decentralized P2P energy markets that coexist with the utility grid in the smart grid of the future. These differences are detailed in the section 3.

### 2.2.5 Peer-to-Peer Energy Trading

Peer-to-Peer (P2P) markets or Transaction Energy (TE) allow small suppliers to compete with traditional providers of goods or services. Their primary function is to make it possible for buyers to find sellers of all scales and engage in *trustworthy and convenient* transactions. P2P energy markets, conceptually, also perform the same function of allowing the *prosumers* to directly share the energy generated on-site with their neighbors or peers.

P2P markets are fundamentally an instrument to decentralize the energy markets as it enables the electricity trade without intermediaries such as a single authority or centralized organization managing the system. However, P2P energy markets have a peculiarity that makes them an interesting use-case of the P2P /decentralization concept - they trade electricity - a commodity which is very difficult to store on large scale and must be available *instantaneously* (on demand). Stocking up electricity *in bulk*[9] is not a pragmatic solution nor it is possible to have customers queue up to get the inventory. Adding to this web of complexities is the intermittent nature of the supply (of renewable generation technologies) as well as continuously varying demand.

Given this unique characteristic of the commodity being traded, P2P *energy market* design is a complex process that must consider the market as well as physical layers of this system, while also accounting for the timescale axis. Furthermore, for the P2P energy markets to co-exist with the incumbent centralized electricity market's physical and virtual infrastructure, their interactions with the wholesale markets must also be considered in designing the framework.

# 3 Proposed Framework: Design and Operational Scheme

## 3.1 A Metaphysical Perspective on Decentralization

Decentralization refers to the transfer of control and decision-making from a centralized entity (individual, organization, or group thereof) to a distributed network. It is not a new concept - in energy systems context or otherwise. In the energy systems context, however, the complexity of this multifaceted concept increases manifolds due to the nature of the commodity being traded (the electricity) and the enormity of the traditional electricity sector's structure that builds, operates, and maintains this capital-intensive network.

While it is tempting to label energy systems in binary terms as *centralized* or *decentralized*, in actuality it is a continuum. The degree to which the energy systems can be decentralized depends on the characteristics of its sub-systems[10], policy implications, and the conditions in the existing energy systems paradigm that either support or suppress the successful deployment of the decentralization concept. Decentralization of the sub-system handling the Information and Communication Technology (ICT), for example, is a process that can be designed and deployed with relatively less friction[11]. Similarly, sub-systems such as market operations, grid management[12], grid infrastructure[13] can be decentralized by analyzing the existing processes in place and the degree to which these existing processes pose resistance towards complete decentralization.

---

[9] The energy storage technologies are available today. But the argument that their scale is sufficiently large to store electricity in bulk such that the majority of generation capacity at any point can be turned off by the ISOs doesn't hold any ground in the present context. A decade or two from now, it is possible that energy storage's falling prices make them a true alternative to utility-scale generation plants, but the research work today can't be hedged with such a risky bet.

[10] examples of sub-systems within energy systems: grid operations, grid infrastructure build-out and maintenance, market operations, etc.

Given the complexities associated with decentralization of energy systems, it is best approached using *systems theory*. United Nations Development Programme aptly defines this approach to decentralization as a "whole systems perspective, including levels, spheres, sectors and functions and seeing the community level as the entry point at which holistic definitions of development goals are from the people themselves and where it is most practical to support them. It involves seeing multi-level frameworks and continuous, synergistic processes of interaction and iteration of cycles as critical for achieving wholeness in a decentralized system and for sustaining its development." [24]. Therefore, in this work, we have taken a *systems* approach to design the framework for the grid's *operations* domain, with the aim of decentralizing the market operations rooted in a prosumer-centric paradigm. Within the operations realm/domain, there are three distinct dimensions that come together to enable operations of P2P energy markets (shown in Figure 2). This work focuses on the design of the decision-making algorithm within the *market sphere*.

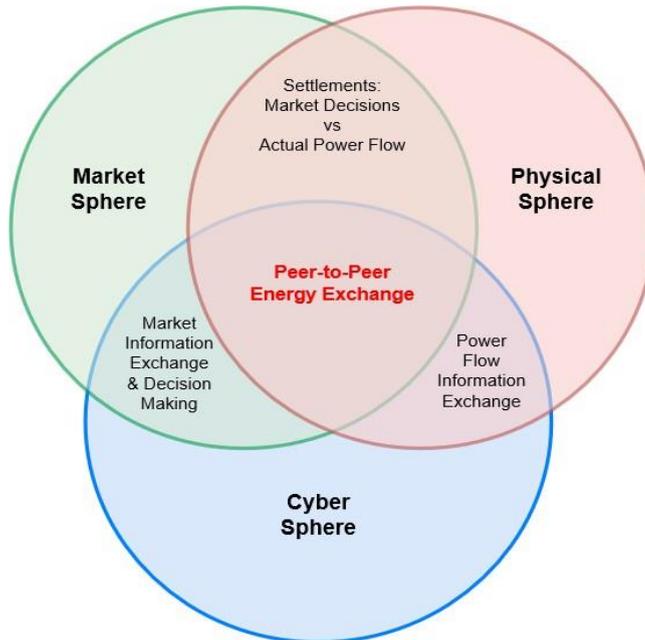

*Figure 2 P2P Energy Markets are actuated by the marriage of three technological spheres.*

In addition to the technical aspects, the social and economic implications of decentralizing energy systems play a crucial role. Embracing decentralization in the energy sector has the potential to foster community empowerment and engagement. Localized energy production and consumption can lead to increased resilience in the face of disruptions, such as extreme weather events or cyber-attacks. Moreover, decentralization can create economic opportunities by encouraging the growth of small and medium-sized enterprises (SMEs) that specialize in renewable energy technologies, energy efficiency solutions, and innovative grid management services. The shift towards a more decentralized energy landscape not only transforms the technical architecture but also redefines the socio-economic dynamics within and beyond the energy sector.

Furthermore, the role of policymakers becomes pivotal in shaping the trajectory of decentralization in energy systems. The regulatory framework must evolve to incentivize and facilitate the integration of decentralized technologies. This includes establishing clear guidelines for the interplay between centralized and decentralized components within the energy grid, as well as addressing issues related to market dynamics, pricing mechanisms, and consumer protection. Striking a balance between fostering innovation and ensuring the reliability and security of the energy supply is a delicate task that requires ongoing collaboration between policymakers, industry stakeholders, and the public. As decentralization continues to reshape the energy landscape, a comprehensive and adaptive regulatory framework will be essential to navigate the evolving challenges and opportunities in this transformative journey.

## 3.2 Existing Challenges and Gaps in the Literature

The growing momentum towards empowering the bottom-up approach to manage and operate the electricity grid calls for inventing approaches to minimize the friction between the long existing top- down and latterly growing bottom-up approaches to grid operations and management. Though there have been a number of works in the literature that tackle peer-to-peer energy market design as well as VPP operations - the decentralization concept has not been dealt with a *systems* approach which considers the incumbents in the industry (i.e. existing wholesale markets, load serving entities, etc.) and position the design of P2P markets such that their inevitable interactions with the central grid (i.e. top-down operations approach) are accounted for. By positioning the design of P2P markets, we mean devising their operational scheme such that the objectives of agents (i.e., prosumers) in the markets are not in disharmony with the objectives of the agents upstream (utilities, load serving entities, aggregators, etc.).

---

[11] can be accomplished using technology like Blockchain [33].
[12] involve economic dispatch and unit commitment considering transmission and distribution grid constraints.
[13] involves building, operating, and maintaining the assets such as generation plants, transmission systems including high voltage lines, transformers, substations, distribution systems, etc.

At the same time, in order to satisfy top-down *economic* constraints[14] that are in place to conduct the globally cost-effective operations of the grid, prosumers shouldn't be kept from operating their DERs to maximize their individual cost savings. In other words, the maximization of prosumer savings, unrestricted by the top-down economic constraints, is the ultimate goal of the bottom-up approach to energy systems operations that we seek to actualize through the proposed framework. It is imperative that given the nature of the electricity sector, this goal is subjected to the physical limits of the underlying power systems infrastructure. The proposition of designing a bottom-up approach such that the primary objective is prosumer benefit maximization while also accounting for the harmonious interactions with the upstream grid operations through ancillary market participation as the secondary objective, to the author's best of our knowledge, has not been put forth in the literature.

Equally important to the effective design of P2P markets is the scaling capabilities of the market framework. As the number of peers in a single marketplace explodes, the performance (speed of convergences of trade, in our use-case) of the system gets negatively impacted. One effective strategy for overcoming this scaling challenge is *partitioning markets* such that local peers are placed in a single module or unit. These modules can then be interconnected in a hierarchical scheme such that different P2P markets are functional at different layers of the hierarchy. These P2P markets (on different layers) are then designed to prudently interact with each other. We elucidate their interaction scheme in the following sub-sections. We employ this tactic of *orchestrated compartmentalization* in our framework to endow it with reasonable scalability potential.

Now, the difficulty in instituting such interconnected hierarchical energy markets is their interoperability in all three technological spheres (physical, market, and cyber as shown in Figure 2). Physical interoperability is, in part, facilitated by the interconnected web of the power distribution lines which provide the physical path for the power to flow between the participants that are part of different layers of the hierarchical markets. Orderly power flow convergence completes the physical interoperability equation. Market interoperability calls for designing P2P trade mechanisms such that it is feasible to conduct *inter and intra* market trades within and between different layers of the hierarchical framework.

Interoperability in the cyber domain, on the other hand, requires the exchange of needed information to solve the power flow as well as market mechanisms to securely enter and store information[15] at various phases of the trading process (bidding, clearing, and settlement) along with providing a smooth interface for information retrieval. The topic of cyber interoperability is out-of-scope for this work.

In addressing the dynamic landscape of decentralized energy systems, it becomes paramount to integrate advanced optimization algorithms and machine learning techniques to enhance the adaptability and efficiency of peer-to-peer (P2P) energy market designs. Leveraging predictive analytics and real-time data processing, the proposed framework aims to synthesize a comprehensive systems approach. This entails deploying intelligent agents within the P2P markets, equipped with autonomous decision-making capabilities, to dynamically adjust to the evolving conditions of the grid.

Simultaneously, within the context of top-down economic constraints, it is imperative to implement sophisticated demand response mechanisms and advanced control strategies to facilitate the optimal utilization of Distributed Energy Resources (DERs) by prosumers. The integration of smart grid technologies, such as real-time demand forecasting and grid state estimation, plays a pivotal role in orchestrating the harmonious interaction between the bottom-up approach and the overarching top-down grid operations. Through the utilization of intelligent agents and advanced control algorithms, the proposed framework ensures that prosumers can harness the full potential of their DERs within the confines of global cost-effective grid operations, thereby achieving a delicate balance between individual cost savings and system-wide efficiency.

## 3.3 Integrated Hierarchical Framework - Empirical Design

The overarching theme behind the proposed design scheme is re-imagining the operations of DERs and the role of prosumers such that the bottom-up approach to electricity grid operations and management is prioritized above the top-down approach yet ensuring that their operational objectives are not diametrically opposite. In the same vein, for addressing the challenges listed in the previous sub-section, our framework is designed to bring together various existing smart grid concepts (DERs, microgrids, VPPs) and synergize their operational objectives to create prosumer-centric P2P energy markets that gel well with the wholesale markets such that prosumers have the option to offer their DERs' services at the wholesale level as well.

Transactive energy (TE) represents a cutting-edge paradigm shift in the smart grid domain, essentially redefining the role of end-users from passive consumers to active participants. This concept, characterized by its broad scope and inclusivity, opens the door to a myriad of interactive possibilities within the smart grid framework. Central to TE is the facilitation of dynamic, two-way energy transactions and information exchanges, allowing end-users not only to consume but also to produce, store, and manage energy in a more integrated and efficient manner. By leveraging advanced technologies like IoT, AI, and blockchain, TE fosters a more decentralized, resilient, and sustainable energy system. This approach significantly enhances grid reliability,

---

optimizes resource utilization, and empowers consumers by providing them with greater control over their energy usage and production, thus catalyzing a fundamental transformation in the energy landscape.

The key to truly decentralized energy systems operations is prosumer-centric decisions — which means having the ability to make decisions with greater autonomy about the use of the energy that's produced or stored on-site. This setup entails that at any point in time, prosumers should have all the choices available to them *simultaneously* (shown in Figure 3).

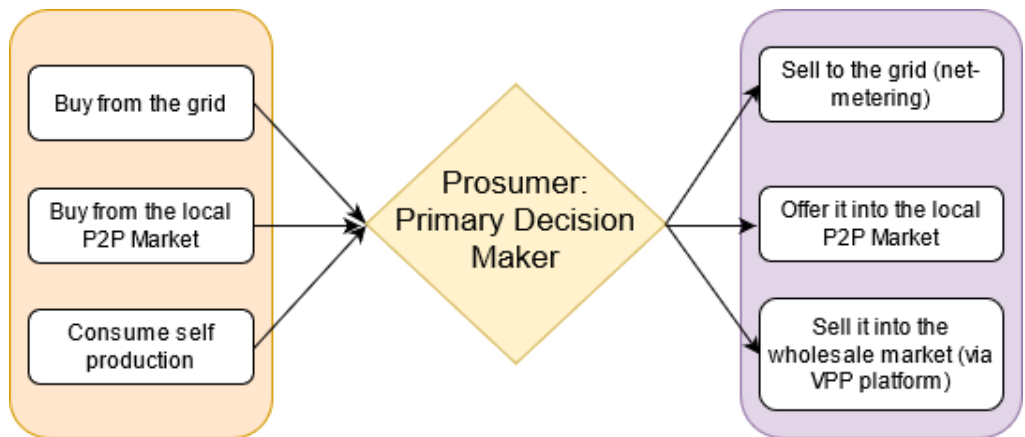

*Figure 3 Prosumer-centric framework: choices to be made available to prosumers*

When the prosumers are free to choose any of the options listed above, they can then plan their operations with the objective of minimizing their energy costs while maintaining their consumption patterns and preferences. For establishing such a prosumer-centric model of the electricity sector's *operational planning*, the existing notion of top-down operational planning needs to be challenged to make space for the bottom-up operational plans to exist. We explain this concept in the following paragraphs.

As per the existing process, the overall power systems planning process starts with the capacity planning (i.e., utility-scale generation, transmission, and distribution infrastructure build-out) for a 3–10-year horizon. *Capacity Markets* mainly operate at this level. The generation entities that get into financially-binding agreements with ISOs to provide generation capacity in the future are compensated through the Capacity Markets. After that, the operational planning starts with looking months into the future to ensure the reliability of the supply while accounting for the major planned outages (centralized power plants or high-voltage transmission lines or transformer maintenance and upgrades). The *bilateral transactions* between large-scale market players, without the ISOs/RTOs acting as mediators, happen within this time horizon (months in advance). In bilateral transactions, the two parties know each other's identities. The *Financial Transmission Rights Market* also operates on this horizon where the transmission capacities are booked to assist the flow of energy that's been committed in the bilateral transaction process.

The *Energy Markets* are then planned and run on a day-ahead (unit commitment) and hour-ahead (economic dispatch) basis. As it gets closer to real-time, *Ancillary Markets* are at play to ensure the reliability of the supply. The most basic requirement of power systems operations – maintaining the supply-demand balance on a micro-second near-real-time basis – can be met either by adjusting the generation or the load. The generation side adjustment mechanisms were explained in section 4. As per the load-side (or demand-side) adjustment mechanisms, the top-down approach gets the prosumers to adjust their consumption patterns in favor of monetary gains. The process is termed as termed a *Demand Response*. The DERs (which also encompass flexible loads) are called upon to decrease their consumption in order to maintain the supply-demand balance. It is here that we propose to make the paradigm shift by raising this question —

> Instead of being driven by top-down operational choices to make adjustments to their own schedules (through economic incentives), why can prosumers not schedule their DERs' dispatch based on their preferences and yet not have to (metaphorically) leave money on the table (which the top-down demand response program's economic incentive would otherwise offer)?

**The above question is at the heart of the bottom-up approach to grid operations.** If we were to raise this question a decade ago, it wouldn't have stood much ground because the future of DERs didn't seem as certain as it seems today. But in the present context, when more and more DERs are being deployed – the decentralization of power systems is arguably *the* optimal path to a low-carbon future. The underlying principle of the decentralization concept is indeed about individuals having choices and also the ability to make their own decisions given those choices. Therefore, we propose that through the mechanisms such as P2P markets and not-for-profit digital aggregators such as VPPs, the operational planning of DERs (owned by prosumers) should be done much like top-down operational planning.

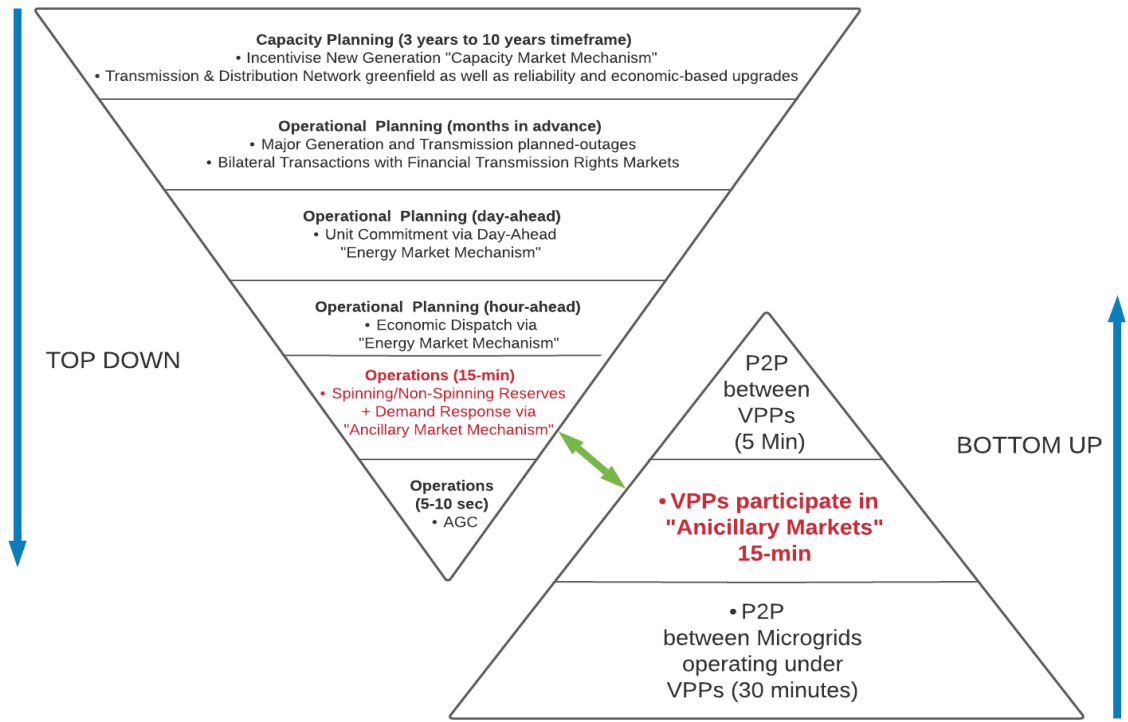

*Figure 4 Marriage of top-down and bottom-up approaches to electricity grid operations*

**However, we do not propose this bottom-up approach as a *replacement* for the top-down approach, we instead offer a coordinated operational planning scheme that helps the two co-exist and coordinate**. Figure 4 depicts the proposed scheme. The inverted first triangle summarizes the existing power systems planning procedure based on the horizon of planning – both infrastructural as well as operational planning. In current distribution systems operations, *the second operational planning triangle is non-existent, which meant DERs would only respond to the demand response signals sent by the utility or the aggregator*. But in our proposed scheme, the prosumers (individual microgrids) get to plan their operations (including the selling or buying energy) well in advance (30-min ahead) and then interact with the demand response programs (non-dispatchable kind) if they find it cost-effective via VPP's aggregation platform (15-min ahead).

The following sub-section describes the above operational scheme at all three layers of the hierarchical framework in detail.

## 3.4 Integrated Hierarchical Framework: Relational Operations Planning

In the proposed framework, the bottom triangle from Figure 1 shows the hierarchy of dispatch decisions that *starts with* the individual prosumers' choices (prosumer choice model depicted in Figure 3). In the Figure 4 Marriage of top-down and bottom-up approaches to electricity grid operations, it is the bottom-most layer of the bottom triangle. Here the prosumers (microgrids in our setup) participate in the 30-minute ahead P2P energy markets operating at the feeder level. The VPPs, acting as a semi-autonomous digital platform for the prosumers to interact, then calculates the overflow from the 30-min ahead P2P markets and offer that aggregated energy (i.e., overflow) into the *Ancillary Markets* that run in 15-min ahead time horizon (the green arrow in the diagram). The third layer in the bottom-up operational planning scheme is P2P energy markets between various VPPs, operating at a 5-min time horizon. This P2P energy market basically acts as a bottom-up *secondary market* for the VPPs. When VPPs are not able to provide the promised energy (in the 15-minute ahead ancillary markets) due to fluctuations in the first layer (P2P between individual microgrids), they conduct bilateral transactions between each other to meet their obligations in the 15-minute ahead Ancillary Markets. Figure 5 depicts the proposed operational planning scheme[16] .

This operational framework, as delineated in Figure 5, ensures a resilient and adaptive distribution grid by integrating energy trading and management at three levels that complement each other with respect to operational timeframe. The system's multi-tiered structure allows for a dynamic balance between supply and demand across different time horizons, thus enhancing the efficiency of the localized and small-scale decentralized markets. The bottom-up approach, starting from individual microgrids and scaling up to VPP interactions, provides a decentralized yet coordinated mechanism for conducting peer to peer trades and selling the spillover in ancillary markets. This is particularly crucial in managing the intermittent nature of renewable energy sources at the microgrid level. The framework's flexibility also enables rapid adjustments to unforeseen demand or supply changes, ensuring reliability and avoiding penalties. Moreover, the incorporation of ancillary markets as a critical component in the hierarchy highlights the framework's ability to optimize energy distribution and maintain grid balance, further demonstrating its robustness in handling the complexities of modern energy systems.

---

[16] The energy *trades*, represented by the dotted lines in the diagram, are different from physical energy *flows*. These flows are dealt with, primarily in the **physical sphere**, while the **market sphere** manages the trades.

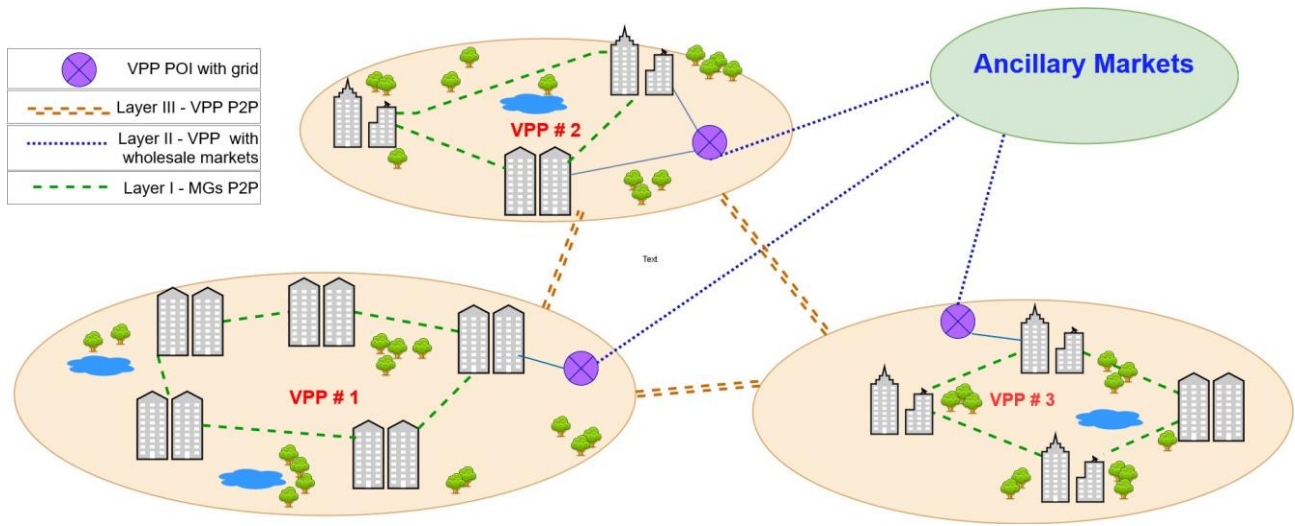

*Figure 5 Hierarchical Framework: the dotted lines represent the energy trades within and among different VPPs. The VPPs are also connected to the wholesale markets on Layer II.*

### 3.4.1 Layer I - Localized Peer to Peer Markets for Individual Microgrids

In our framework, individual microgrids are the *behind-the-meter* resources which means the energy consumption and injection is measured at the point of interconnection between the distribution grid and the microgrid's main meter. The DERs assets within the boundaries of the microgrid site are operated and dispatched at the prosumer/owner level. This means that the objective of the intelligent microgrid controller is to minimize the energy cost by taking all the external price information[17] and internal consumption and preferences[18] as input to the optimization problem.

These microgrids (we call them prosumers hereby) are the participants of the P2P energy markets at Layer I (bottom-most layer in the inverted triangle of Figure 5). Planning their dispatch 30-min ahead, they make buy or sell offers in the P2P energy markets that are digitally conducted in the cyber sphere. Each feeder, with multiple prosumers, has a P2P market conducted using the *bilateral contracts mechanism*. It is important to note that in the United States, the length of a feeder in power distribution networks can vary significantly. A three-phase feeder main can be fairly short, in the order of a mile or two, or it can be as long as 30 miles. The length of feeders is closely linked with load density at the location. For instance, in an area where the customer load density is strong, the primary network will end very close to consumers and secondary feeders will be short [25].

The steps involved in conducting such organized P2P energy trades are summarized in Figure 6. This arrangement is an involved interplay of sequential processes that span over all three spheres of energy systems operations (physical, cyber, and market). It starts with prosumers interacting with the market sphere to place bids (for buy or sell) and then the cyber sphere recording the data[19]. One notable feature in the process flow (from Figure 6) is that the P2P market sphere in Layer I leads the overall P2P trading mechanism. Once the energy flow at a specific point in time has occurred, the measurements are then read and recorded in the cyber network. As the last step, settlements are done after the fact.

The third step is power flow which is conducted to ensure the reliability of the power flow in the feeder and upstream network. Power flow is conducted by the autonomous agent that constitutes the VPP and is not a part of the individual prosumer's microgrid control scheme. The microgrid controller ensures the voltage and frequency stability of the MG at the Point of Interconnection (POI) through its hierarchical control scheme.

The power flow analysis, on the other hand, is serving as an auxiliary process that has the objective of ensuring that any reliability concerns are not overlooked (for example, persistent voltage violations) within the feeder. The results from the power flow can raise red flags which can then be used by the VPP platform (explained in the Layer II subsection below) to send signals to the prosumers to increase/reduce the real and reactive power they are injecting into the distribution grid's feeder. At what price a prosumer wants to buy/sell their energy is not determined by the power flow, however. This kind of arrangement between the market sphere and the physical sphere can be afforded in this setup because of the small-scale nature of the energy flow and transactions on the feeder level that forms Layer I P2P energy markets in our proposed framework.

---

[17] utility energy price, P2P market buy/sell prices, compensation for committing energy to VPP for ancillary market.

[18] flexible and inflexible consumption predictions, battery state of charge, etc.

[19] The detailed modeling of the cyber-sphere, aimed at ensuring data security and prosumer privacy, is out of scope for this work. However, it is worth noting that blockchain technology is a promising candidate for designing such a robust cyber sphere.

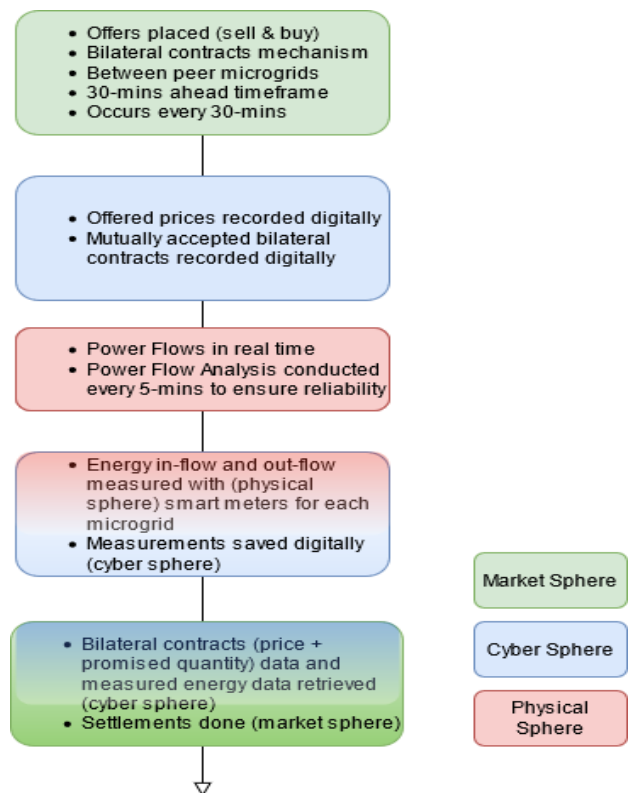

*Figure 6 Layer I - P2P Energy Markets Design and Operations*

Another noteworthy feature of Layer I P2P Market design is that it is operating largely independent of the upstream market operations. That is to say, it is truly a grass-root approach where the front-line prosumers are having the opportunity to make their energy production and consumption decisions without any direct influence of the top-down wholesale or retail market operations[20]. And so, it follows that the price determination mechanism for Layer I P2P energy markets, which can be reasonably assumed to be free markets[21], can be effectively done using a combination of *model predictive control* and the *game theoretic approach*.

## 3.4.2 Layer II - Participation in Ancillary Markets: Feeder Level Virtual Power Plants

The second layer of our proposed framework is essentially serving as a bridge between the top-down and bottom-up market operations. It offers the mechanism for the individual prosumers to participate in the wholesale markets, thereby joining the sub-set of market participants who are 'price-setters' - which is one of the core qualities of the decentralization paradigm. The decision-makers in Layer II are VPPs. VPPs, which we defined as semi-autonomous non-profit digital platforms, are designed with the operational objective of maximizing the energy cost savings of Layer I participants (a note on ownership structure of VPPs[22]).

As per the functionality, Layer II is primarily managing the over-and/or-under spill of the energy from Layer I P2P energy markets. It then utilizes the overflow of energy to make bids in the 15-min ancillary services markets. The operations of the VPP platform's market and cyber sphere are highly intertwined because VPP is constantly monitoring the status of Layer I P2P markets, based on which it decides the amount and price of the bids it places into the ancillary markets. It also participated in the physical sphere of the operations by housing and running the power flow program. While running the power flow, if any reliability issues are flagged by the outcome, VPP communicates this information to the participants of the Layer I P2P markets and suggests corrective measures. The functionality of the VPP platform is summarized in the flow chart below (Figure 7).

The choice of time horizons for Layer I (30-min) and Layer II (15-min) in this framework are noteworthy. Layer I operate with 30-minute ahead bids for increments of 30-min time slots. Layer II operates with 15-min ahead window for increments of 15-minute time slots such that it can make re-adjustments to its bidding strategy in the 15-minute ancillary markets based on the measured under-or-over spill of the energy from Layer I. In this process, Layer II is assisting Layer I participants (i.e., individual prosumers) extract more value for their energy by selling it in the wholesale (ancillary) markets. The time horizon of this layer, thus, is shorter than Layer I.

---

[20] We specify it to be *direct* influence because indirectly the price point of the Layer I P2P market will be influenced by the prices upstream because prosumers do have the choice of either buying the energy from the utility or from P2P market. So, this competitive market setup will have second and third-order causalities. However, the determination of such causalities is out of the scope of this work.

[21] By definition, the free market is an economic system in which prices are determined by unrestricted competition between privately owned businesses (prosumers in our use-case).

[22] The capital cost of setting up VPPs digital infrastructure is a Layer I community expense. By making it a *community-owned* platform, VPPs are not serving as third-party aggregators trying to maximize their own profits by having Layer I participants sign up for financially binding dispatchable and non-dispatchable demand response programs. Instead, they are serving as an intelligent **semi-autonomous agent working for** the Layer I community to represent them in the wholesale markets by performing the aggregation of the assets owned by the prosumers in the community.

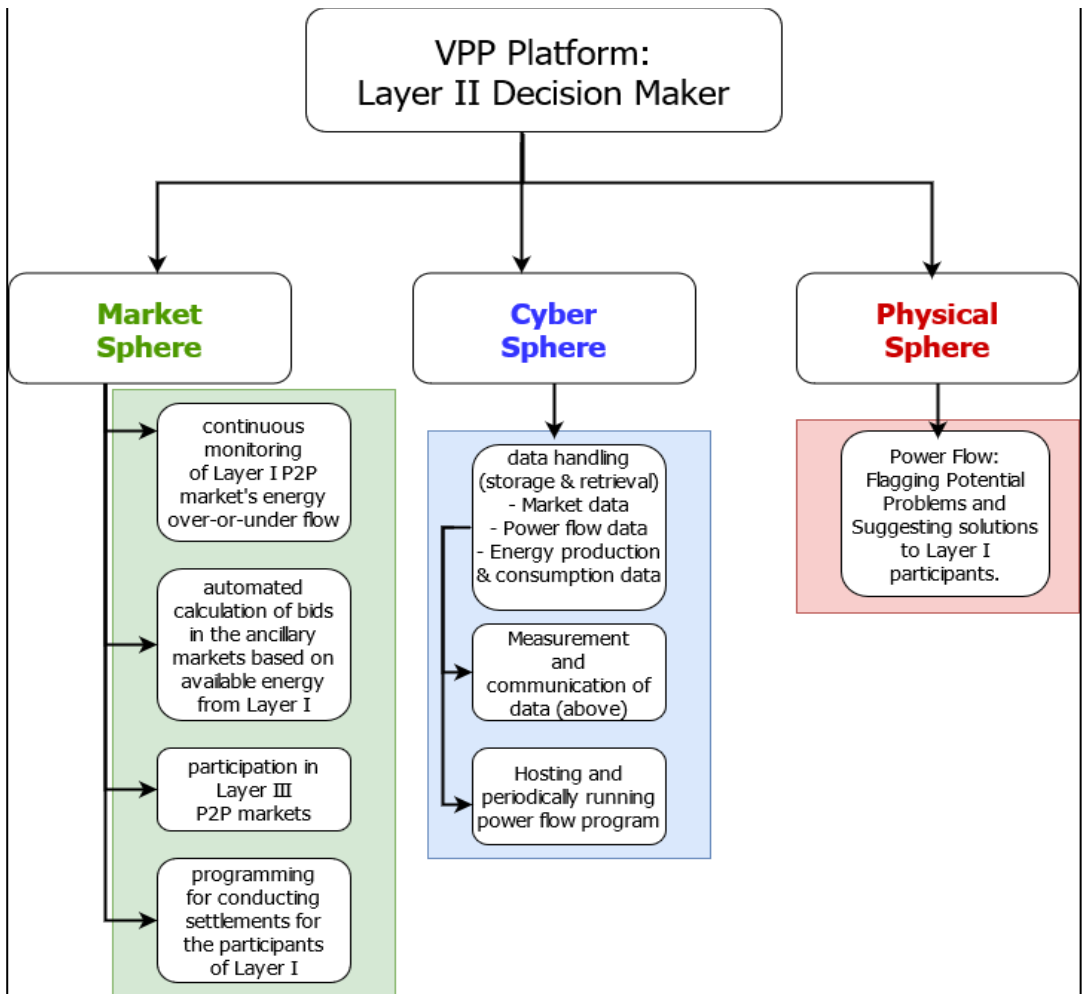

*Figure 7 Functionalities of the semi-automated VPP platform - Layer II Decision Maker*

### 3.4.3 Layer III - Peer to Peer Markets for VPPs

Layer III operates P2P energy markets between the VPPs, with the shortest time horizon (5-minute). The objective of P2P markets at Layer III is to manage the under-and-over spill from Layer II. That is to say, if a VPP commits a certain amount of energy in the ancillary market for a 15-minute time slot (which is 15-min into the future) and later finds out that the under spill from the Layer I P2P energy market will make it impossible for the VPP to deliver the committed energy in the ancillary market, it can then procure the needed amount of energy by participating in the Layer III P2P market. In other words, the Layer III P2P market is designed to be the *secondary market* for the VPPs to buy energy in 5-minute increments so that the penalty charges by the ancillary markets can be avoided.

Figure 8 tabulates the operations of the three layers of the framework.

| Tasks | Time Scale | Operational Objectives | Functionalities | Ownership Structure |
|---|---|---|---|---|
| Layer I (Microgrids P2P Market) | 30-mins ahead (in 30-min increment) | • Cost-minimization<br>• Maintain Prosumer Preference | • MG internal energy management<br>• Intelligent cyber layer designed to find optimal buy-sell decisions | All the assets (DERs, flexible loads, energy storage) are owned by the behind-the-meter prosumer |
| Layer II (VPP /aggregators participate in Ancillary Markets) | 15-mins ahead (in 15-min increment) | • Manage the "overflow" of net positive or net negative energy from the Layer I<br>• Offer the extra energy into the ancillary markets | • Simultaneously work with Layer I and Layer III with the objective of maximizing profits of Layer I participants | May or may not own a "community battery" |
| Layer III (VPP P2P Markets) | 5-mins ahead (in 5-min increment) | • Manage the over-or-under flow from Layer II through bilateral contract mechanism | • Serve as "secondary markets" for VPPs to buy or sell energy to meet the promised quantity in the ancillary markets | No asset ownership |

*Figure 8 Hierarchical P2P Energy Markets - Operational Details of the Three Layers.*

## 3.5   Merits of the Proposed Framework

The framework is designed by considering the bottom-up operational scheme's timescales to ensure that the decision-making hierarchy indeed flows from the bottom most layer. The operation of Layer I is scheduled 30-minutes in advance because the primary objective is to ensure that prosumers can make their own operational decisions without having to necessarily follow the top-down signals of modifying their usage (for example, demand response signal sent with 5-min advance notice). The VPP, functioning as a non-profit semi-autonomous mediator platform, operates at 15-min timescale so that it can accommodate the over and under flow of the energy from the layer (where P2P energy markets operate with 30-min timescale). Layer III of the framework, on the other hand, operates with a 5-minute timescale because its function is to assist VPPs in meeting their energy needs for the bids they have placed in the ancillary markets - thereby, accommodating the over and underflow of the energy from Layer II operations.

Though the design of the framework is primarily optimized for prosumer preferences and cost savings, its utility is enhanced because of its compatibility with the existing distribution grid operations which are traditionally rooted in the top-down approach. The framework encompasses the majority of the smart grid constituents (microgrids, DERs, VPPs, etc.) and meticulously re-frames their objectives and functionalities to bring these technologies together to serve the vision of bottom-up smart grid operations. Moreover, it offers an effective mechanism to get DERs to participate in wholesale markets, making FERC Order 2222's execution relatively frictionless. The salient features of the framework can be summarized as follows:

- On a meta-level, our framework is offering a credible path forward to holistically decentralize energy systems operations - this is an effective answer to the question of energy equality.
- Designs comprehensive decentralized operational paradigm - seemingly diversified smart grid technologies like DERs, microgrids, VPPs are brought together to work in tandem for creating a truly prosumer-centric bottom-up approach to energy systems operations.
- Delineates the differences between and the roles of the market, physical, and cyber spheres – within the context of *P2P market operations domain*. And maps out the convoluted interactions between these spheres for establishing effective P2P energy market mechanisms.
- Efficacious in coordinating centralized and decentralized energy markets - utilizing VPPs as the connecting link between the two.
- Instead of acting as a third-party aggregator, VPPs are modeled as semi-autonomous non- profit entities. Thereby, effectively benefiting the prosumers.
- Scalable: The hierarchical framework equips the proposed P2P energy markets to scale efficiently.
- Interoperable[23]: The interoperability of different layers is designed to overcome the potential timescale conflicts that would arise in the operational planning of different layers in the hierarchy.

---

[23] Scalability and interoperability are two important characteristics that determine the real-life deployment potential of the proposed decentralized P2P energy markets design

# 4 Integrated Hierarchical Framework: Layer-wise Technical Modeling

In this section, we present the technical modeling of the proposed framework, which consists of three layers: the microgrid layer, the virtual power plant (VPP) layer, and the peer-to-peer (P2P) layer. Each layer has its own objectives, constraints, and decision variables, which are formulated as optimization problems and game theoretic problems. The interactions between the layers are also modeled using appropriate coordination mechanisms. The technical modeling aims to capture the physical and market aspects of the framework, as well as the interplay between the top-down and bottom-up approaches to grid operations. The following subsections describe the technical modeling of each layer in detail.

## 4.1 Layer I – Microgrid Controls and Game Theoretic P2P Trades

The implementation of microgrid controls in Layer I - Localized Peer to Peer (P2P) Markets is a crucial aspect of our proposed framework. In this P2P market which is conducted within the confines of Layer 1 participants, individual microgrids operate as autonomous entities, each equipped with advanced control system. These microgrid controllers are designed to optimize the local energy balance, taking into account the dynamic nature of Distributed Energy Resources (DERs) and the fluctuating energy demands within the microgrid.

At the core of the control system is an optimization algorithm that continuously monitors and predicts energy production and consumption patterns. We employ Model Predictive Control (MPC) for this purpose [26]. This algorithm not only considers the current state of the microgrid but also integrates forecasts of renewable energy generation (like solar and battery storage) and also integrates the forecasted consumption within the individual microgrids. By doing so, it can make real-time decisions on energy dispatch, storage, and exchanges with other microgrids in the P2P market. The control system also plays a pivotal role in the bidding process within the P2P energy market. It calculates optimal bids for buying or selling energy based on the microgrid's current energy surplus or deficit, market prices, and predefined user preferences. This enables each microgrid to act as a smart and proactive market participant, optimizing its financial returns while maintaining a stable and efficient energy supply.

Participating in the P2P trading process, the objective function of MPC is to maximize profits from the energy trades. The objective function involves revenue from selling energy, costs from buying energy, and penalties for imbalances or deviations from scheduled trades. Mathematically, it could be represented as:

$$J = \sum_{t=0}^{N-1} (p_{sell}(t)) \cdot (E_{sell}(t) - \left(p_{buy}(t)\right) \cdot (E_{buy}(t)))$$

Where $p_{sell}(t)$ and $p_{buy}(t)$ are selling and buying prices at time $(t)$, and $E_{sell}(t)$ and $E_{buy}(t)$ are the corresponding energy quantities. The MPC framework involves solving an optimization problem at each time step. This problem aims to find the optimal bidding strategy (energy quantities to buy or sell) over a prediction horizon considering the forecasted data and the objective function. The optimization is subject to constraints such as:

- Energy balance constraints (demand must be met either by generation, storage, or purchases).
- Generation and storage limits.
- Network constraints (maintaining voltage and frequency at POI).
- Market rules (e.g., minimum or maximum bid sizes, bidding time windows).

MPC operates on a rolling horizon basis. At each time step, it uses the latest available data to update its forecasts and re-optimizes the bidding strategy for the upcoming period. This approach allows the microgrid to adapt to real-time changes in market conditions, demand, or generation capacity. The pseudo code for MPC modeling is given below:

```
MPC_Optimization_For_Microgrid_Prosumer():
    Initialize:
        T = Time horizon for MPC (12 hours)
        dt = Time step (30 minutes)
        Energy_Buy_Price_P2P = Local P2P energy market buying prices [array of length T]
        Energy_Buy_Price_Utility = Utility tariff prices for buying energy from the central grid [array of length T]
        Energy_Sell_Price_NetMetering = Net-metering prices set by utility [array of length T]
        Energy_Sell_Price_P2P = Local P2P energy market selling prices [array of length T]
        Demand_Forecast = Forecasted energy demand [array of length T]
        Generation_Forecast = Forecasted energy generation from microgrid [array of length T]
        Energy_Storage_Capacity = Maximum storage capacity of energy storage system
        Energy_Storage_State = Initial state of charge of energy storage system
        Energy_Storage_Charge_Discharge_Rate = Max charge/discharge rate
    For each time step t in T:
        Define Decision Variables:
            Energy_To_Sell_P2P[t] = Amount of energy to sell in P2P market at time t
            Energy_To_Sell_Utility[t] = Amount of energy to sell to utility at time t
            Energy_To_Buy_P2P[t] = Amount of energy to buy from P2P market at time t
            Energy_To_Buy_Utility[t] = Amount of energy to buy from utility at time t
            Energy_To_Charge_Storage[t] = Amount of energy to charge storage at time t
            Energy_To_Discharge_Storage[t] = Amount of energy to discharge from storage at time t
        Define Constraints:
            - Balance energy supply and demand at each time step
            - Ensure energy transactions do not exceed generation or demand
            - Enforce storage capacity limits and charge/discharge rates
            - Other technical or regulatory constraints
        Define Objective Function:
            - Maximize revenue from selling energy (considering both P2P and net-metering prices)
            - Minimize cost of buying energy (considering both P2P and utility prices)
            - Include other objectives such as minimizing grid reliance, maximizing renewable usage

        Solve Optimization Problem:
            - Use an appropriate solver to maximize the objective function subject to the defined constraints

        Update Energy_Storage_State based on charging/discharging decisions

    Return:
        Optimal energy bidding, selling, buying, and storage decisions for each time step in T
```

Figure 9 shows the hierarchical control scheme which is employed by prosumers (individual microgrid participants of Layer I P2P markets) for energy management and bidding in the Layer I P2P energy market at the feeder level.

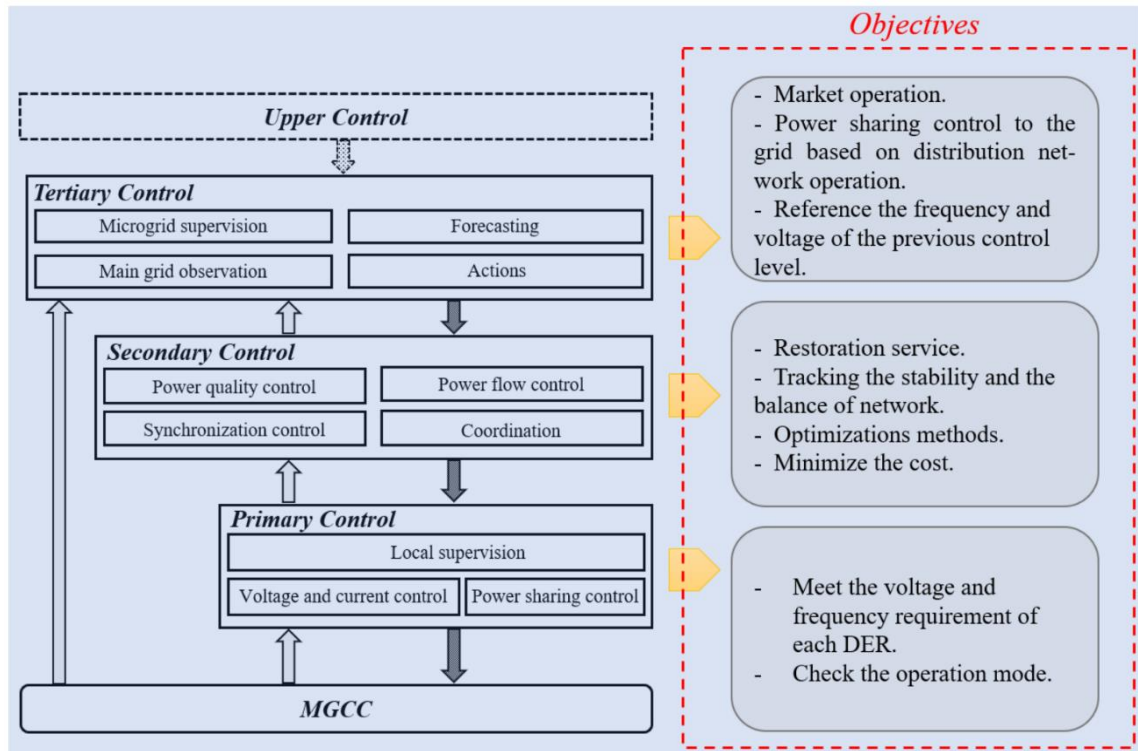

*Figure 9 Microgrid Hierarchical Control Scheme [27]*

Another important function of microgrid controls in Layer I is ensuring that it meets the DER integration standard at the Point of Interconnection (POI). The standards are detailed in Appendix IV. These systems are equipped with capabilities to detect and mitigate potential issues such as voltage fluctuations or frequency deviations. In the event of grid disturbances or failures in the main grid, the microgrid control system can swiftly isolate the microgrid from the main grid (islanding) and manage the local generation and load to keep the microgrid operational. Additionally, these control systems facilitate a higher level of consumer participation and empowerment. Residents and businesses within the microgrid can interact with the microgrid controller to set preferences for energy consumption, generation, and storage. This not only enhances the efficiency of the microgrid but also allows consumers to play an active role in the energy market.

### 4.1.1 Non-cooperative game theoretic approach for price determination

Peer-to-peer (P2P) energy trading represents a transformative approach in the energy market, characterized by decentralized transactions. The application of game theory in this domain primarily focuses on designing strategies for participants to optimize their benefits while ensuring a stable and efficient energy market [28]. Game theory models in P2P energy trading usually involve multiple players, including energy producers, consumers, and prosumers (who both produce and consume energy), each with their own utility functions. These models aim to predict the behaviors of these players in response to various market conditions, tariffs, and regulations [29]. For example, in a non-cooperative game scenario, each participant aims to maximize their own payoff without considering the overall welfare of the system. This approach can lead to equilibrium states where market dynamics stabilize, albeit potentially at the cost of reduced overall efficiency or fairness.

On the other hand, cooperative game theory models in P2P energy trading emphasize collective strategies that aim to optimize the welfare of the entire network [30]. These models typically involve the formation of coalitions among participants, allowing for more effective management of energy distribution and consumption, and often resulting in better pricing mechanisms and grid stability. Such cooperative strategies are particularly relevant in scenarios where the energy grid is strained or in the presence of significant renewable energy integration. In our framework, we employ a non-cooperative game theory model. Non-cooperative Game Theory, particularly in P2P energy trading, involves strategizing in a competitive market where each participant aims to optimize their own payoff. When these methods are applied together, MPC can be used to optimize the bidding or trading strategy of an individual player, taking into account the likely actions of others as predicted by game-theoretic models.

In this approach, each player has a set of strategies they can choose from. For prosumers with interest in selling, this might involve setting different price levels for different time-of-day for their access energy production; for buyers, strategies could involve choosing at what price they buy the energy to maximize their savings (e.g. prosumers with battery storage can choose to buy during cheaper rates and use it when prices are high in the Layer I P2P energy market). Each player's strategy leads to a certain payoff, which depends on the strategies chosen by other players. In price determination, the payoff could be profit for sellers or utility for buyers. A payoff matrix is used to represent the payoffs for each combination of strategies chosen by the players. Players choose their strategies based on their best response to the strategies of others. A Nash equilibrium is reached when each player's strategy is the best response to the strategies of the others, meaning no player has anything to gain by changing only their own strategy unilaterally. In the context of price determination, the Nash equilibrium would represent a set of prices where sellers are maximizing their profits given the prices set by other sellers and the demand of the buyers, and buyers are maximizing their utility given the available prices and their preferences.

## 4.2 Layer II – VPP, modeled as an autonomous agent

A Virtual Power Plant (VPP) operating as an autonomous agent (all digital) presents an innovative approach that our framework proposes. We model it as an all-digital non-profit entity. As a non-profit entity, this *digital construct is uniquely owned* by participants across multiple Peer-to-Peer (P2P) energy markets operating in Layer I, each corresponding to different feeders within the first layer (Layer I) of an intricate energy framework. This design enables a decentralized yet cohesive management of Layer I P2P markets, leveraging the collective power of individual prosumers - those who both produce and consume energy. The VPP, by summing the access production from all the P2P Layer I market, acts as a single entity that represents a set of feeders (each feeder running its dedicated P2P energy market for multiple microgrid participants within that feeder).

The core objective of this VPP is to strategically sell excess energy, referred to as the 'overspill' from Layer I, to the wholesale market realm, specifically targeting the ancillary market. This overspill is essentially the surplus energy generated by the Layer I participants, who are the prosumers. The VPP harnesses this excess, which might otherwise be sold at a low "net-metering" price to the utility, and channels it into the *lucrative* ancillary market. The aim is to achieve the highest possible price for this energy, thereby maximizing cost-savings for the participants of Layer I P2P energy markets who are also the owner of the all-digital autonomous agentic VPP of Layer II.

### 4.2.1 MPC as algorithm powering the autonomous agent

MPC is well-suited for modeling and managing the operations of a Virtual Power Plant (VPP) as described above. The primary strength of MPC lies in its ability to anticipate future conditions and optimize control actions accordingly, which is critical in the dynamic and complex environment of a VPP. Given that the VPP acts as a collective representation of multiple Peer-to-Peer (P2P) energy markets, each with its unique energy production and consumption patterns, MPC's predictive capabilities enable it to efficiently balance and forecast the energy needs and overspill of these diverse markets.

Moreover, MPC's ability to handle multiple constraints and objectives simultaneously aligns perfectly with the VPP's goal of maximizing the price of excess energy sold to the wholesale market and furthermore participate in a secondary P2P energy market in Layer III of the proposed framework. This optimization is crucial as it directly impacts the cost-savings for the participants who are also the owners of the VPP. By predicting market trends, energy demands, and supply fluctuations, MPC can strategically determine the best times to sell the overspill energy, thus ensuring the highest possible revenue. In essence, MPC's foresight, adaptability, and multi-objective optimization capabilities make it an ideal tool for the VPP's mission to efficiently and profitably manage energy resources in a complex and fluctuating ancillary market environment.

## 4.3 Layer III – Secondary P2P market for VPPs

The purpose of P2P markets at Layer III is to balance the spillage from Layer II. These markets operate P2P energy transactions between the VPPs, with the shortest time horizon (5-minute). If a VPP pledges a certain amount of energy in the ancillary market for a 15-minute time slot (which is 15-min ahead) and later realizes that the spillage from the Layer I P2P energy market will prevent the VPP from delivering the pledged energy in the ancillary market, it can then acquire the required amount of energy by participating in the Layer III P2P market. In other words, the Layer III P2P market is designed to be the secondary market for the VPPs to sell energy in 5-minute increments so that the penalty charges by the ancillary markets can be avoided.

### 4.3.1 Cooperative game theoretic approach

We propose to conduct the P2P trading amongst VPPs in Layer III using a cooperative game theoretic approach. Incorporating cooperative game theory into the operation of Layer III P2P energy markets can enhance the efficiency and reliability of Virtual Power Plants (VPPs) in conducting secondary energy transactions. This approach can be implemented through the following technical mechanisms [31]. VPPs form coalitions based on their energy requirements and availability. By employing cooperative game theory, these coalitions can be dynamically adjusted in real-time, considering the fluctuating energy demands and supply in the 5-minute time horizon of Layer III. This allows VPPs to optimize their strategies for buying or selling energy, leading to more efficient market outcomes. Fair distribution of the gain or costs among the participating VPPs can be achieved using Shapley Value Allocation. Furthermore, cooperative game theory offers various bargaining solutions that can be employed in Layer III markets. These solutions help in determining the optimal prices for energy trades among VPPs, taking into account the individual utility functions of each VPP and the overall market efficiency. Also, dynamic pricing models can be developed that adapt to the real-time supply-demand dynamics of the Layer III market. These models can help individual VPPs in making more informed decisions about when to buy or sell energy, optimizing their costs and revenues. Cooperative game theory can also aid in managing network congestion, which is crucial in the short-term 5-minute market. By predicting and responding to congestion patterns, VPPs can adjust their strategies to maintain grid stability and avoid additional charges. Implementing game-theoretic approaches allows for the development of risk mitigation strategies, as VPPs can analyze potential outcomes and their probabilities. This helps in making decisions that minimize risks associated with energy commitments and ancillary market volatility.

By integrating these cooperative game-theoretic approaches, Layer III P2P energy markets can operate more efficiently, allowing VPPs to effectively manage energy transactions in the secondary market. This not only improves the financial performance of VPPs but also adds to the choices that are available to the Layer I prosumers ultimately leading to the maximized profit for energy they sell. The pseudo code for implementation of the cooperative game theory approach is as follows:

```
Algorithm: Non-Cooperative Game Theory for P2P Energy Trading among VPPs
Inputs:
- Energy requirements and availability for each VPP
- Historical and predictive data on energy demands and supply
- Ancillary market constraints and penalty charges
- Time slots for energy trading (5-minute increments)
Procedure:
1. Initialize VPP Network:
   - For each VPP, input initial energy availability and requirements.
   - Establish network connections between VPPs for P2P trading.
2. Run Real-Time Market Dynamics:
   - Every 5 minutes, update the energy requirements and availability for each VPP.
   - Adjust the VPPs' strategies based on real-time data.
3. Form Coalitions Dynamically:
   - Based on current energy needs and supplies, dynamically form coalitions among VPPs.
   - Utilize cooperative game theory principles to optimize coalition formation.
4. Calculate Shapley Value for Payoff Distribution:
   - For each coalition, compute the Shapley value to fairly allocate payoffs among VPPs.
   - Ensure equitable distribution based on each VPP's contribution to the coalition.
5. Implement Bargaining Solutions:
   - Negotiate energy prices within coalitions, considering individual utility functions and market efficiency.
   - Employ game-theoretic bargaining solutions to determine optimal prices.
6. Apply Dynamic Pricing Models:
   - Develop and apply pricing models that adapt to real-time market dynamics.
   - Assist VPPs in decision-making for buying or selling energy optimally.
7. Manage Network Congestion:
   - Predict and respond to congestion patterns in the 5-minute market.
   - Adjust trading strategies to maintain grid stability and avoid penalties.
8. Develop Risk Mitigation Strategies:
   - Analyze potential market outcomes and their probabilities.
   - Make decisions that minimize risks associated with energy commitments and market volatility.
9. Trade Energy in P2P Market:
   - Conduct energy transactions based on the established strategies, prices, and coalition agreements.
   - Continuously monitor and adjust strategies based on market developments and VPP objectives.
10. Evaluate Market Performance:
    - After each trading cycle, evaluate the performance of each VPP and the overall market.
    - Use feedback to refine strategies and improve future market operations.
Output:
- Completed energy transactions for each VPP
- Updated energy availability and requirements for the next cycle
- Performance metrics for market efficiency and VPP profitability
```

## 4.4. Comprehensive Transaction Process – An illustrative example

To bring all the pieces of the framework described in this article together, we present a flowchart demonstrating the process in Figure 10. This framework is comprised of three distinct layers, each with specific functions and objectives:

- **Layer I** - Localized Peer to Peer Markets for Individual Microgrids: This layer facilitates peer-to-peer energy markets at the individual microgrid level, allowing prosumers to actively participate in energy trading within their localized networks.. It also allows prosumers to have simultaneous choices of buying and selling energy from utility grid as well as local P2P energy markets. Moreover, and perhaps most importantly, interaction of Layer I with Layer II offers prosumers the choice to participate in wholesale market's ancillary markets.
- **Layer II** - Participation in Ancillary Markets: Feeder Level Virtual Power Plants: Layer II acts as a bridge between the top-down and bottom-up market operations. It enables individual prosumers to participate in wholesale markets, positioning them as 'price-setters' in the market. This layer is managed by Virtual Power Plants (VPPs), which are semi-autonomous, non-profit digital platforms. Their primary goal is to maximize the energy cost savings for participants in Layer I.
- **Layer III** - Peer to Peer Markets for VPPs: This layer operates P2P energy markets between different VPPs and has the shortest time horizon of 5 minutes. The primary function of this layer is to manage the surplus or deficit of energy from Layer II. For instance, if a VPP commits to delivering a certain amount of energy in the ancillary market but later realizes a shortfall due to under-delivery from Layer I, it can then purchase the required energy through the Layer III P2P market. This helps in avoiding penalties in the ancillary markets.

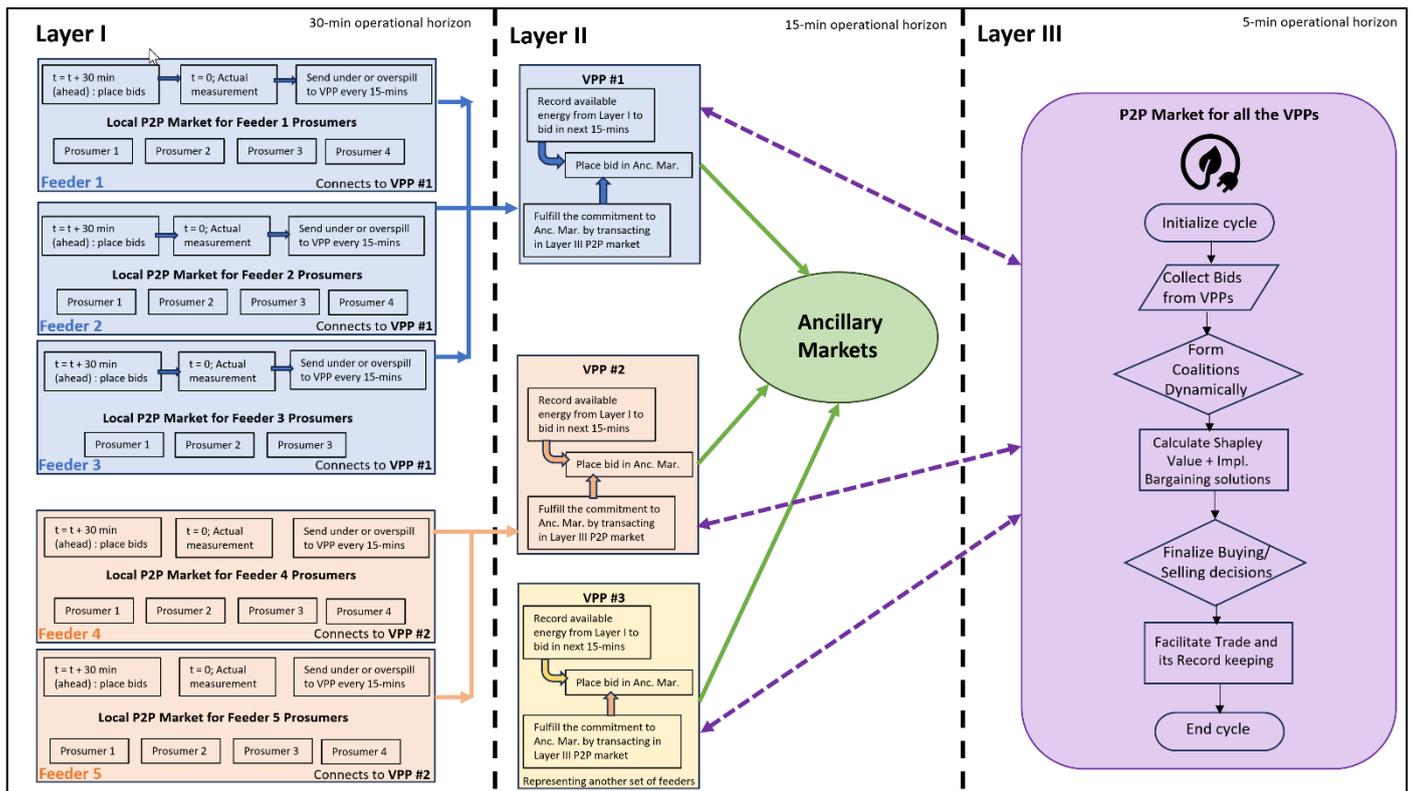

*Figure 10 Process flow diagram of the proposed framework*

The overarching theme of this framework is to reimagine the operations of DERs and the role of prosumers, integrating various smart grid concepts like DERs, microgrids, and VPPs. This design aims to create prosumer-centric P2P energy markets that work in tandem with wholesale markets, thereby transforming end-users from passive consumers to active participants in the energy markets – on both local and wholesale market spheres simultaneously.

# 5 Conclusions and Outlook

Traditionally designed for unidirectional flow of power from centralized power plants, the grid now experiences bidirectional flow as the power generated by DERs is injected back into the grid. With DER deployment, consumers, who were considered passive users of electricity, are becoming 'prosumers'. They undertake proactive behavior by managing their consumption, production, and energy storage. The DERs are the primary fuel for the growing *localized generation* trend. These DERs are considered behind-the-meter (BTM) sources; the deployment of BTM DERs is primarily driven by individual commercial, industrial, and residential customers rather than the utilities or regulators. Localized generation, also termed as DERs, is advantageous in multiple ways for fulfilling the objectives of the transforming energy system including i) Preventing losses from the long-distance transfer of power; ii) Integration of more clean energy on the grid (e.g. rooftop solar) resulting in reduced carbon footprint; iii) Increased resilience - ability to sustain site's critical load during utility grid outages; iv) Reduction in upstream *generation & transmission* capacity requirements (which is a considerable capital expense for both the aging grid in developed nations and the grid which is yet to build in the developing nations); v) Decentralization of the electricity production and supply mechanism decreases the vulnerability of society towards utility grid outages, which is, otherwise the only source of power for the majority of the population.

Though localized generation undoubtedly has numerous benefits, incentivizing end-users to invest in the deployment of DERs requires their individual benefits to outweigh the costs. Moreover, the traditional utility revenue generation models do not gel well with the DERs' concept of selling power back to the grid. To sustain their operations going forward with the widespread DER deployment trends, utilities have begun to restructure their business models that previously relied on the traditional unidirectional energy flow value chain. Along with these financial and economic considerations, from a technical perspective, the uncertain and distributed nature of these DERs poses challenges to reliable and efficient grid operations. This is because, in the past, the grid operations mechanisms were designed for centralized large-scale power plants to generate electricity and send it over a high-voltage transmission network. The design was suitable for maintaining the reliability of the electricity supply using fossil-fuel-powered generation plants to meet the uncertain demand. However, the inherent intermittency and uncertainty associated with renewable energy based DERs render the traditional operation mechanism less efficient.

Nevertheless, owing to the numerous benefits of localized renewable energy generation, it is worthwhile to solve the aforementioned challenges for facilitating their effective integration with the central grid. Hence, the way forward is to develop and implement both business and technical operations models that harness the capabilities of DERs' and complement their shortcomings. And that is where the contribution of this work lies - developing a framework for bottom-up energy systems operations that gels well with the existing top-down operational paradigm. The framework focuses on harnessing the synergies between the two approaches to grid-management by orchestrating the time-horizons of dispatch-decisions of different agents in the network. The framework is aimed at developing the operational scheme that places the prosumer's decision-making at the center while ensuring that the centralized grid-operators do get benefit in the process (when VPPs participate in the ancillary services markets).

The paper's framework aims to design a prosumer-centric grid operations model that integrates the top-down and bottom-up approaches of electricity market operations and management in favor of prosumer's savings and preferences. The practical applications of this integration are:

- It allows prosumers to have more choices and autonomy over the use of their distributed energy resources (DERs), such as selling, storing, or consuming the energy they produce or store on-site.
- It enables prosumers to simultaneously participate in both local peer-to-peer (P2P) energy markets and centralized wholesale markets, creating multiple value streams and revenue opportunities for them. Even more importantly, the proposed mechanism ensures that, given the choice to participate in both local and centralized markets, prosumers choose the one which is more lucrative, thereby operating as an autonomous/algorithmic/digital agent that represents prosumers unbiasedly in the process.
- In this way, it fosters the true decentralization and democratization of the energy sector where prosumers don't have to be price-takers of utility's decisions (example: net-metering scheme) or depend on a third-party VPP provider to try and sell the energy that they generate. In this way, our proposed framework empowers the bottom-up actors and reduces their dependence (as price-takers) on large-scale utilities and intermediaries.

## 5.1 Future Works

There are three distinct yet closely connected operational layers in this framework where bottom-up P2P market dispatch decisions take place. This work can be taken forward in the following directions (Layer-wise categorization):

- Layer I: Taking the case study forward by incorporating the scenario where the prosumers have the additional choice of committing their energy to the VPPs for selling in the ancillary services market.
- Layer II: The semi-autonomously non-profit VPP platform's decision-making can be simulated using an optimization problem where the objective of the VPP is to maximize prosumer income through selling their energy in the wholesale market (ancillary services) while obeying the resource constraints posed by the individual prosumer choices that are passed down from Layer I (the bottom most layer in the second triangle of Figure 3.1).
- Layer II: The cost-benefit analysis for adding a community energy storage on Layer II, managed by the VPP, can be conducted to assess the potential profits that VPP platform (acting as non-profit digital mediator) can gain by bidding the stored energy into the ancillary markets (along with the energy that's being offered by the prosumers from the Layer I).
- Layer III: Analyzing the operations of multiple VPPs, that are physically connected by distribution lines, and are conducting a separate P2P market that runs at the shortest time horizon (5-min) with the aim of making up for the shortage or access of the energy supply that individual VPPs have (after bidding in the ancillary markets with 15-min time horizon).
- Layer III: Analyzing the operations of multiple VPPs P2P energy markets (the case above) without direct physical connections (i.e., distribution lines) connected all the VPPs to each other.

All the future research directions (lying within the market sphere from 3.1) described above are heading towards the goal of quantitatively establishing the holistic robustness of the proposed framework by comparing it against the existing top-down approach - the indicator of supremacy being the **prosumer-savings without having to trade off their preferences** such as consumption patterns and information privacy.

Another interesting pathway to direct future research is within the cyber sphere. Implementation of this framework requires the exchange of critical infrastructure information as well as the information related to prosumer's public and private bidding decisions that drive the P2P energy market's price points. It also involves establishing intelligent and autonomous decision-making modules that reside on the servers (or on peer nodes in the case of serverless digital infrastructure). Moreover, the exchange of this information needs to be secure and fast enough to keep up with the market dynamics. The challenge of transmitting this information securely such that agents (who are authorized) to access the information are able to do so seamlessly is an information technology-centered conundrum. Devising mechanisms and apparatuses to facilitate digital communication, storage, and processing is an interesting research direction to explore further using Blockchain technology.

sector: A systematic review of challenges and opportunities," *Renewable and Sustainable Energy Reviews,* vol. 100, pp. 143-174, 2019.
[34] M. Uddin, M. F. Romlie, M. F. Abdullah, S. A. Halim, A. H. A. Bakar and T. C. Kwang, "A review on peak load shaving strategies," *Renewable and Sustainable Energy Reviews,* vol. 82, no. 3, pp. 3323-3332, 2018.

# Appendix I

**Summary of factors driving the transformation of electricity sector globally is given below:**

- Climate change (extreme heat and cold) - resulting in the pressing need to reduce carbon footprint [31]
- Ever-growing energy needs of the planet driven by population growth, urbanization, and eco- nomic development of the developing nations. Need to find a way to "sustainable-ly" meet this energy demand.
- Aging infrastructure of the central grids in the developed countries - needing major capital investments for upgrades to maintain the reliability of the supply.
- Rise in low-probability high-impact events such as natural disasters (hurricanes, earthquakes, etc.) calling for increased resilience of the power grid to sustain the critical loads in the face of such events.
- Technological advances such as:
    - Large-scale Distributed Energy Resources (DER) deployment - driven by increased scales of production and falling costs - leading to higher levels of adoption.
    - Proliferation of consumer-level information and communications technologies - leading to the Internet-of-Things (IoT) age of hyper-connectivity.
    - Wide-spread deployment of measurement and sensing infrastructure- due to decreasing sensor costs - leading to a plethora of data availability.
    - Growing number of electric vehicles on the roads - even higher adoption predicted in the near future.
    - Increase in *grid-responsive high-performance* buildings equipped with advanced analytics and controls.

# Appendix II

**Whos' Who in the Electricity Sector - *Agents* in Energy Systems:**

Designing effective market operations in the energy sector requires in-depth consideration of the organization and roles of different players or agents that frequently interact in the process of bringing energy from generation to the end-users.

- **Electric utilities** are public or private entities that own and operate the transmission and distribution lines that transport electricity and supply it to households and businesses.
- **Vertically integrated utilities** are electric utilities that own not only transmission and distribution infrastructure but also the power plants responsible for electricity generation. They control every stage of the electricity production process in the areas they serve.
- **Load serving entities** (or local distribution companies) deliver electric power to customers. In some regions, load serving entities generate their own electricity, while in others they simply provide the grid infrastructure required to deliver electricity to customers.
- **Electricity retailers** sell electricity to consumers in regions with deregulated electricity markets. In these regions, **local distribution companies** own the distribution infrastructure and deliver electricity to customers.
- **Grid operators** (also known as **transmission system operators**) balance grid operations by ensuring that the amount of electricity put into the grid matches the amount of electricity used by consumers. They work with all of the utilities, generators, and retailers to ensure that the grid is balanced and reliable: too little power can cause blackouts, while too much can cause damage to equipment.
- In some regions, electric utilities act as grid operators. In other areas, the grid is operated by **regional transmission organizations** (RTOs) or **independent system operators** (ISOs), which are different types of organizations that operate the grid but do not own the resources and infrastructure (such as power plants and power lines) within it. Following deregulation, RTOs replaced utilities as grid operators and became the operators of wholesale markets for electricity. These RTOs have evolved over time.

# Appendix III

# Wholesale Markets and Demand Response

The three types of market mechanisms within wholesale energy markets in the United States are as follows:

## Energy Markets

Energy markets are auctions that are used to coordinate the production of electricity on a day-to-day basis. In an energy market, electric suppliers offer to sell the electricity that their power plants generate for a particular bid price, while load-serving entities (the demand side) bid for that electricity in order to meet their customers energy demand. Supply-side quantities and bids are ordered in ascending order of offer price. The market clears when the amount of electricity offered matches the amount demanded, and generators receive this market price per megawatt hour of power generated. Prior to selecting the least cost set of resources needed to meet the expected energy demand on a daily basis, ISO evaluates factors such as weather forecasts, generator availability, transmission line outages, etc. Energy markets are further divided into *day ahead* and *real-time* energy markets.

The day-ahead market schedules electricity production, ancillary services commitments, and consumption before the operating day. The day-ahead energy market produces *financially binding schedules* for the production and consumption of electricity one day before its production and use (the operating day). In day-ahead markets, the schedules for supply and usage of energy are compiled hours ahead of the beginning of the operating day. The RTO/ISO then runs a computerized market model that matches buyers and sellers throughout the market footprint for each hour throughout the day. The model then evaluates the bids and offers of the participants, based on the power flows needed to move the electricity throughout the grid from generators to consumers. Additionally, the model must account for changing system capabilities that occur, based on weather and equipment outages, plus the rules and procedures that are used to ensure system reliability. Generation and demand bids that are scheduled in the day-ahead market are settled at the day-ahead market prices.

The real-time market reconciles any differences between the schedule in the day-ahead market and the real-time load while observing reliability criteria, such as forced or unplanned outages and the electricity flow limits on transmission lines. The real-time market is run in five-minute intervals and clears a much smaller volume of energy and ancillary services than the day-ahead market. The real-time market also provides supply resources with additional opportunities for offering energy into the market. When the real-time generation and load are different from the day-ahead cleared amount, the difference is settled at the real-time price. Real-time market prices are significantly more volatile than day-ahead market prices.

## Capacity Markets

Electricity retailers are required by the North American Electric Reliability Corporation (NERC), an independent organization that ensures grid reliability, to support enough generating capacity to meet forecasted load plus a reserve margin to maintain grid reliability. Some RTOs run a capacity auction to provide retailers with a way to procure their capacity requirements while also enabling generators to recover fixed costs, i.e. those costs that do not vary with electricity production, that may not be covered in the energy markets alone. Capacity markets are meant to provide financial incentives for suppliers to keep generation assets online and to induce new investment in generation. Capacity markets are generally forward markets to have generation capacity online and are ready to produce electricity at least one year ahead of time.

The capacity market provides incentives that stimulate investments in the maintenance of the existing generation as well as the development of the new sources of capacity to ensure the future reliability of the power system. The capacity market auction works as follows: generators set their bid price at an amount equal to the cost of keeping their plant available to operate if needed. Similar to the energy market, these bids are arranged from lowest to highest. Once the bids reach the required quantity that all the retailers collectively must acquire in order to adequately meet expected peak demand plus a reserve margin, the market clears, or supply meets demand. At this point, generators that cleared the market, or were chosen to provide capacity, all receive the same clearing price which is determined by the bid price of the last generator used to meet demand. Capacity markets provide two fold benefits: i) They provide a mechanism for ensuring longer term reliability by securing enough power to meet the needs of electricity consumers a few years (3 as an example) in the future and ii) The payments to generators in the capacity market are essentially a reward for that generator being available to operate and provide electricity if needed, ensuring medium-term reliability of the supply. Consequently, if generators are unavailable to operate during a time when they are called upon, they may face fees under capacity performance requirements.

## Ancillary Markets

Ancillary Services Markets allow the ISO/RTO to maintain a portfolio of backup generation in case of unexpectedly high demand or if contingencies, such as generator outages or a transformer failure, arise in the system. This way ancillary markets support both *supply reliability* and *transmission* of the electricity. These services are produced and consumed in real-time, or in the very near term.

There are many different types of ancillary services, corresponding to the speed with which the backup generation needs to be dispatched. The minimum amount of each ancillary service required to maintain the grid reliability are established by NERC and regional entities. They can be divided into two major types: i) Regulation and ii) Operating Reserves. Regulation matches <u>generation</u> <u>with very short-term changes</u> in load by moving the output of the selected resources up and down via an *automatic control signal*, typically every few seconds. The changes to output are designed to maintain system frequency at 60 Hertz. Operating reserves, on the other hand, are additional ancillary services that are needed to restore load and generation balance when a <u>supply resource</u> <u>trips offline</u>. Operating reserves are further classified into i) Spinning; ii) Non-spinning; and iii) Supplemental reserves based on the speed with which they can start supplying electricity when the dispatch signals are released.

# Wholesale Market Pricing

Owing to the deregulated nature of the electricity markets, most of the wholesale markets today rely upon competitive market forces to set prices[24]. In this research work, our focus is on studying the interactions of the *distribution grid level coordinated DERs* with the wholesale markets based power systems operations (In the U.S., FERC Order 2222 has enabled DERs to participate along- side traditional resources in the regional organized wholesale markets through aggregations [20] ). In competitive markets, the factors driving supply and demand are reflected in the prices. The supply side of the electricity wholesale market equation incorporates both generation and transmission. Both these components must function adequately to meet the demand from all the end-users simultaneously, instantaneously, and reliably.

Consequently, key supply factors like fuel costs, capital costs, transmission capacity and associated constraints, and the operating characteristics of power plants affect the prices. So do the changes on the demand side. In real-time operations, demand is changing all the time. And when the DERs are added to the mix of demand-side changes, they further add to the uncertain nature of the electricity demand. ISOs have established the concept of Locational Marginal Pricing (LMP) to account for the factors described above.

## Locational Marginal Pricing

Locational marginal pricing is a way for wholesale electric energy prices to reflect the value of electric energy at different locations, accounting for the patterns of load, generation, and the physical limits of the transmission system. The RTO/ISO markets calculate an LMP at each location on the power grid. The LMP reflects the marginal cost of serving load at the specific location, given the set of generators that are being dispatched and the limitations of the transmission system. LMP has three elements: an energy charge, a congestion charge and a charge for transmission system energy losses. LMPs are calculated every five minutes in majority of the ISOs. The LMP at a load zone is a weighted average of all the nodes within the load zone.

If the system were entirely unconstrained and had no losses, all LMPs would be the same, reflecting only the cost of serving the next increment of load. The generator with the lowest-cost energy offer available would serve that incremental megawatt of load, and electric energy from that generator would be able to flow to any node on the transmission system. LMPs differ generally among locations because transmission and reserve constraints prevent the next-cheapest megawatt (MW) of electric energy from reaching all locations of the grid. Even during periods when the cheapest megawatt can reach all locations, the marginal cost of physical losses will result in different LMPs across the system.

## "Price-taker" vs "Price-maker" Nodes in Wholesale Markets

The generation units and the Load Serving Entities (LSE) that bid in the market for selling and buying energy respectively, are considered the price-maker nodes. Both these types of firms have a say in the market clearing price that ISO/RTO calculates while finding the point of intersection of the supply and the demand curve. On the other hand, DERs that reside behind a customer meter are not able to participate in the wholesale market and hence are considered price-takers. That is, a market participant that is not able to dictate the prices in the market. Price-taker models, in the power systems, a called "behind-the-meter" resources. These limitations of DERs are discussed in detail in the following sections.

The electricity bill that gets generated for the end-user (considered largely "price-taker" nodes) has a number of components that determine the total cost. These price components vary with different region and operators. The three major constituents for the price of electricity are: i) cost of generating electricity; ii) cost of transmitting electricity over long distances using a high-voltage transmission network; and iii) cost of distributing electricity by the retailers using the low-voltage distribution network.

## Demand Response

Electricity demand is generally insensitive to price, meaning that demand does not typically fall when prices rise. This occurs for several reasons, including that most end-use consumers of electricity are not exposed to real-time electricity prices. However, some utilities and grid operators have developed ways to stimulate a response from consumers through demand-response programs. Demand response (DR) is a short-term, voluntary decrease in electrical consumption by end-use customers that is generally triggered by compromised grid reliability or high wholesale market prices. In exchange for conducting (and sometimes just committing) to curtail their load, customers are remunerated. Figure 11 aptly summarizes the types of demand response programs.

---

[24] Also, note that not all the territories in the region or country have wholesale markets operated by the Independent System Operators either. Canada, for example, has some provinces like Manitoba which are operating on the traditional vertically integrated structure where large monopoly provides of bundled electricity services dominate the market. Prices in such an arrangement are set based on the service providers cost of service. On the other hand, Alberta's electricity sector is based on market competition.

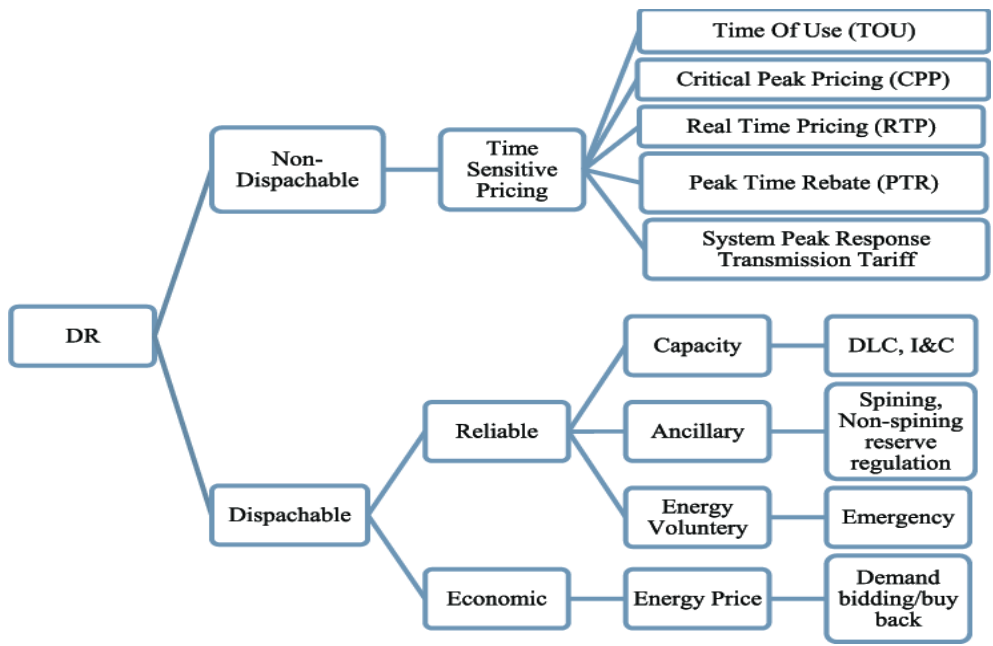

*Figure 11 Classification of demand response programs [22]*

Different types of demand response programs are administered by both retail and wholesale entities including market operators, utilities, load serving entities, third-party aggregators. In this work, we focus on the non-dispatchable type of demand response because the primary objective is to empower the prosumers to make their energy *use and trade decisions* (under peer-to-peer market setup, while also staying connected to the central grid).

# Appendix IV

**IEEE Standards for Microgrid Controls**:
IEEE standards provide a framework for research and industry practices in Energy Management and Control Systems (EMCS). They guide researchers to work within a common framework and compare new methodologies with standardized benchmarks. Following are the main IEE standards associated with EMCS:

1. **IEEE STD. 1547**: Standard for Interconnecting Distributed Resources with Electric Power Systems:
- Focuses on technical specifications and testing requirements for interconnection and interoperability between utility electric power systems and Distributed Energy Resources (DERs) up to 10 MVA.
- Covers microgrid transition issues and outlines voltage and frequency regulation limits for 60Hz grids.
- Includes ride-through capability requirements for DERs under voltage and frequency deviation.

2. **IEEE STD. 1547.3**: Guide for Monitoring Information Exchange and Control of Distributed Resources Interconnected with Electric Power Systems:
- Facilitates interoperability of DERs.
- Supports technical and business operations, including monitoring information exchange and control.
- Emphasizes data exchange performance, security, and communication infrastructure.

3. IEEE STD. 1547.6: Recommended Practice for Interconnecting Distributed Resources with Electric Power Systems Distribution Secondary Networks:
- Provides guidelines for DER interconnection on distribution secondary networks.
- Focuses on technical issues related to islanding, restoration strategies, and network protection.

4. IEEE STD. 2030: Guide for Control and Automation Installations Applied to the Electric Power Infrastructure:
- Offers best practices and approaches for achieving smart grid interoperability.
- Defines integrated architectural perspectives and data flow characteristics for interoperability.
- Focuses on smart grid research, including resilience against cyber-attacks and performance of cyber-enabled control schemes.

5. IEEE STD. 2030.6: IEEE Approved Draft Guide for the Benefit Evaluation of Electric Power Grid Customer Demand

- Provides a guide for demand response applications in smart grids.
- Establishes methods for DR baseline calculation and comprehensive benefit evaluation.
- Highlights energy-sharing frameworks, optimization of load scheduling, and financial savings in demand side management.

6. IEEE STD 2030.7: IEEE Standard for the Specification of Microgrid Controllers:
- Addresses the need for uniform criteria and requirements for microgrid controllers.
- Defines functional requirements for dispatch and transition functions.
- Emphasizes the importance of microgrid self-healing capabilities, substation control, and efficient power management strategies.